\documentclass[aps,twocolumn,nofootinbib,preprintnumbers,groupedaddress]{revtex4}

\usepackage[dvips]{color}
\usepackage[normalem]{ulem}
\usepackage{amsmath}
\usepackage{enumerate}
\usepackage{amsfonts}
\usepackage{epsfig}
\usepackage{yfonts}

\usepackage{graphicx}
\usepackage{bm}

\newcommand\comment[1]{}

\newcommand\om{\omega}
\newcommand\Om{\Omega}
\newcommand\ov{\over }

\def\le{\left}
\def\ri{\right}

\def\({\left(}
\def\){\right)}
\def\[{\left[}
\def\]{\right]}
\def\<{\langle}
\def\>{\rangle}
\def\Dslash{\rlap{\hskip0.2em/}D}

\def\tr{\mathop{\rm tr}}

\newcommand\half{{\ensuremath{\frac{1}{2}}}}

\newcommand\field[1]{{\ensuremath{\mathbb{{#1}}}}}

\newcommand{\RR}{\field{R}}

\newcommand{\ZZ}{\field{Z}}

\newcommand{\be}{\begin{equation}}
\newcommand{\ee}{\end{equation}}
\newcommand{\bea}{\begin{eqnarray}}
\newcommand{\eea}{\end{eqnarray}}
\newcommand{\bwt}{\begin{widetext}}
\newcommand{\ewt}{\end{widetext}}

\newcommand{\bi}{\begin{itemize}}
\newcommand{\ei}{\end{itemize}}
\newcommand{\ben}{\begin{enumerate}}
\newcommand{\een}{\end{enumerate}}
\newcommand{\bca}{\begin{cases}}
\newcommand{\eca}{\end{cases}}
\newcommand{\bln}{\begin{align}}
\newcommand{\eln}{\end{align}}
\newcommand{\bst}{\begin{split}}
\newcommand{\est}{\end{split}}

\newcommand\sR{{\ensuremath{{\mathcal R}}}}

\renewcommand{\Im}{\textrm{Im}\,}
\renewcommand{\Re}{\textrm{Re}\,}

\def\Dslash{\rlap{\hskip0.2em/}D}

\def\spinor{\Psi}
\def\spinors{\psi}

\def\mspinor{m}  %_\zeta}

\begin{document}

%

%\title{Marginal Fermi liquids from holography}

\title{Holographic Fermi surfaces near quantum phase transitions}

\author{David Vegh}
\affiliation{\it Simons Center for Geometry and Physics, Stony Brook University, Stony Brook, NY 11794-3636}

\date{\today}
%
%\date{June 09, 2010}
%\hskip.2in
\begin{abstract}

We study holographic Fermi surfaces coupled to a bosonic degree of freedom using the gauge/gravity correspondence.
The gravity background is a charged black hole in asymptotically AdS spacetime. We introduce a neutral scalar field with parameters such that the system is at a quantum phase transition point.
We further introduce a  Dirac field and couple it to the scalar. At finite $N$, these fields interact in the bulk.
We compute the bulk one-loop contribution to the boundary fermionic self-energies at various quantum phase transition points.
The results would give an embedding of the marginal Fermi liquid model by Varma et al. into a holographic context. We comment on important issues that arise at low energies when one wants to take the one-loop results more seriously.

\end{abstract}

\maketitle

\section{Introduction}

A theoretical understanding of high-temperature superconductors \cite{Bednorz} is one of the major challenges in physics.
Cuprate superconductors exhibit many unconventional properties. They have a generic phase diagram parametrized by the electron (or hole) doping $x$, and the temperature $T$.
The superconducting phase covers a dome-shaped region in the phase diagram. The material is said to be optimally doped where the phase-transition temperature $T_c(x)$ is maximal.

Above the superconducting dome ($T>T_c$) lies the so-called strange metal phase. In this phase, the material has unusual transport properties, e.g. the resistivity has a robust linear temperature dependence up to very high temperatures. In order to fit experimental data, Varma et al. proposed a phenomenological scale-invariant ansatz for the spin and charge fluctuations,
\be
  \label{eqn:fluc}
  \Im \chi(\om) \sim
\left\{ \begin{array}{ll}
  -\chi_0  { \omega \over T} & {\rm for} \quad {|\om| \ll  T} \\
-\chi_0 {\rm sign}\, \om & {\rm for} \quad {T \ll |\om| \ll \om_c}
\end{array} \right.
\ee
where $\om_c$ is a UV cutoff (a few times smaller than the Fermi energy). The fermion self-energy at one-loop is
\be
  \Sigma(\om) \sim \om \log {x \ov \om_c} - i {\pi \ov 2} x
\ee
where $x = \max(|\om|, T)$. This self-energy gives qualitatively different behavior compared to the Fermi liquid case where $\Im \Sigma \sim \om^2$. Near the Fermi surface, the quasiparticle peak is much broader and the quasiparticle weight vanishes logarithmically.
This system was dubbed the {\it marginal Fermi liquid} (MFL).

In recent years, the gauge/gravity correspondence \cite{Maldacena:1997re, Gubser:1998bc, Witten:1998qj} has been applied to describe strongly coupled quantum field theories at finite charge density. The gauge theory is dual to a gravitational theory with negative cosmological constant. The gravity background has an ``ultraviolet'' boundary where it approaches anti-de~Sitter space. ``Infrared'' boundaries may be introduced by black hole horizons. Dynamics near the horizon is generically related to low-energy phenomena in the dual gauge theory.
Recently, holographic Fermi surfaces have been observed in phenomenological (``bottom-up'') models \cite{Lee:2008xf, Liu:2009dm, Cubrovic:2009ye, Faulkner:2009wj, Denef:2009kn, Faulkner:2010tq} (see \cite{Gauntlett:2011mf, Belliard:2011qq} for negative results using a ``top-down'' approach). For certain regions in the parameter space, these  Fermi surfaces are non-Fermi liquids due to their strong dissipation into a locally quantum critical system which is described by the near-horizon $AdS_2\times \RR^n$ spacetime \cite{Faulkner:2009wj, Faulkner:2010tq}.
Marginal Fermi liquids can be obtained by tuning the parameters of the bulk spinor field \cite{Faulkner:2009wj}.
{\it A main question is whether marginal Fermi liquids can arise in a more robust way without tuning the fermion parameters.} See \cite{Jensen:2011su} for a related semi-holographic work.

\begin{figure}[h]
\begin{center}
\includegraphics[scale=0.65]{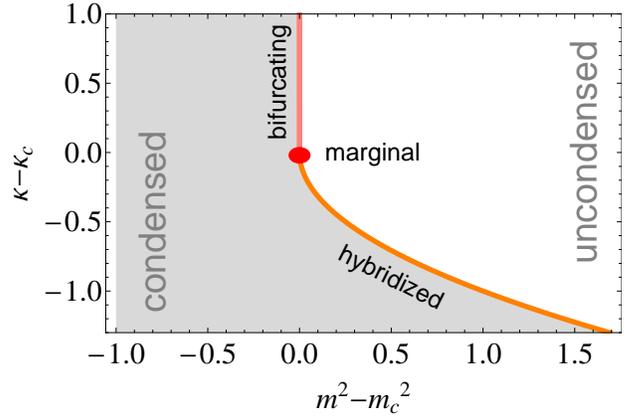}
\caption{\label{fig:phase} Local phase diagram parametrized by the scalar mass $m^2$ and the double trace deformation $\kappa$. At the intersection of the bifurcating and hybridized quantum critical lines lies the marginal critical point.}
\end{center}
\end{figure}

In this paper, we study the holographic system near ``magnetic'' quantum phase transitions. The local structure of the phase diagram is seen in FIG. \ref{fig:phase}.
The magnetic order parameter is described by a neutral scalar field \cite{Iqbal:2010eh}. The boundary response function of the scalar takes the form of (\ref{eqn:fluc}) for the MFL critical point and -- up to logarithmic corrections -- for the bifurcating critical line. We then introduce a bulk Dirac field and couple it to the scalar. At finite $N$, the boson interacts with the fermion.
Fermion self-energy computations reduce to ordinary Feynman diagram calculations with effective vertices which are determined by radial integrals in the bulk. Assuming that the self-energy contributions from other fields (graviton, gauge field) are less relevant, the one-loop fermionic self-energy will have a marginal Fermi liquid form at extremely low energies. Finally, on the hybridized critical line we obtain marginal Fermi liquid and non-Fermi liquid type contributions to the boundary fermionic self-energies, depending on the spacetime dimensions.

The paper is organized as follows. In Section II, we describe the gravity background, the phase diagram and the two-point functions of the fields. Section III discusses how the computation of the  bulk loops can be reduced to a field theory calculation. Using these results, in Section IV we compute the one-loop self-energy diagrams.
We comment on the remaining important questions in the discussion section.

\section{Ingredients}

In this section, we briefly review the basic ingredients of the holographic system. Let us consider a (2+1)-dimensional relativistic conformal field theory (CFT) with a gravity dual and a global $U(1)$ symmetry. We turn on a finite chemical potential for the $U(1)$, and put the system at finite temperature. According to the gauge/gravity duality, this {\it boundary} field theory can be described by the {\it bulk} geometry: a black hole in (3+1)-dimensional anti-de Sitter (AdS$_4$) spacetime. Fermion fields in the bulk are dual to (composite) fermionic operators on the boundary. The current in the CFT is mapped to a $U(1)$ gauge field in the bulk. We further consider a neutral scalar field in the bulk. The expectation value of the dual operator will be a ``magnetic'' order parameter.

\subsection{Black-hole geometry}

The action for a vector field $A_M$ coupled to AdS$_4$ gravity can be written as
 \be \label{grac}
 S = {1 \ov 2 \kappa^2} \int d^{4} x \,
 \sqrt{-g} \le[\sR + {6 \ov R^2} - {R^2 \ov g_F^2} F_{MN} F^{MN} \ri]
 \ee
where $g_F^2$ is an effective dimensionless gauge coupling. The equations of motion
are solved by~\cite{Romans:1991nq, Chamblin:1999tk}
\be
ds^2  =  {r^2 \ov R^2} (-f dt^2 + d \vec x^2)  +  {R^2 dr^2 \ov r^2 f}
\ee
\be
f = 1 + { Q^2 \ov r^{4}} - {M \ov r^3}, \quad A_0 = \mu \le(1- {r_0 \ov  r}\ri), \quad
\mu  \equiv  {g_F Q \ov R^2 r_0}
\ee
This geometry is a charged black hole in $AdS_4$ with horizon located at $r_0$ which is determined by $f(r_0) =0$. Using the AdS/CFT prescription, $\mu $ is identified as the chemical potential of the boundary theory.
In the zero-temperature limit, the near-horizon metric becomes $AdS_2 \times \RR^{2}$.
For simplicity, we rescale the coordinates,
 \bea
&& r \to r_0 r, \quad (t,\vec x) \to {R^2 \ov r_0} (t , \vec x), \quad A_0 \to {r_0 \ov R^2} A_0, \nonumber \\
&& \qquad\quad M \to M r_0^3, \quad Q \to Q r_0^{2}
\eea
to arrive at the metric
\be \label{bhmetric}
{ds}^2 = r^2  (-f dt^2 + d\vec x^2)  + {1 \ov r^2} {dr^2 \ov f}, \qquad
\ee
The horizon is now at $r=1$ and we have
\be \label{eep}
f = 1 +{Q^2 \ov r^{4}} - {1+Q^2 \ov r^3}, \quad A_0 = \mu \le(1- {1 \ov r} \ri), \quad
\mu  = g_F Q  \ .
\ee
The dimensionless temperature is given by $T =  {3-Q^2 \ov 4 \pi}$.
In the zero-temperature limit, the near-horizon metric becomes $AdS_2 \times \RR^{2}$ where the curvature radius of $AdS_2$ is $ R_2 = {R / \sqrt{6}} $. We will use $\zeta$ to denote the radial coordinate in $AdS_2$. At small temperatures, the near-horizon geometry is approximately a black hole $AdS_2\times \RR^{2}$,
\be
  ds^2  \sim  { 1- {\zeta^2 / \zeta_0^2 } \ov \zeta^2} d\tau^2 + {d\zeta^2 \ov \zeta^2\le( 1- {\zeta^2 / \zeta_0^2 }\ri)}  + d\vec x^2
\ee
where $\zeta_0 = (2\pi \tilde T)^{-1}$ and $\tilde T$ is the temperature corresponding to the $\tau$ time coordinate.

\subsection{Matter fields}

We would like to adjust the parameters such that the systems is near a quantum critical point. The magnetic phase will be characterized by a neutral order parameter. The dual bulk field is a scalar with action,
\be
  S_b =  \half \int d^{4} x \,\sqrt{-g} \le( \nabla_M \phi \nabla^M \phi  - V(\phi)  \ri)
\ee
with $  V(\phi) = {1 \ov 4R^2}  \le(\phi^2 + {m^2 R^2} \ri)^2
- {m^4 R^2 \ov 4 } $.
Classically, the scalar will be set to zero. %Nevertheless, its presence affects the other bulk fields.

Let us now introduce a Dirac field, $\spinor$  with $U(1)$ charge $q$.
The fermionic action takes the form,
\be
  %S_\textrm{probe} =
  S_f= i \int d^{4} x \,\sqrt{-g}    \overline{\spinor}  \left( \Gamma^M \overleftrightarrow
  D_M - \mspinor +   \eta  \phi \right) \spinor    + h.c.
\ee
where  $\eta$ parametrizes the coupling of the fermions to the magnetic order parameter.

\subsection{Fermionic Green's function}

 We are going to use the following basis for the gamma matrices,
\be
\Gamma^{\underline{r}}=
\begin{pmatrix}
-\sigma^3  &   0 \\
0 & -\sigma^3
\end{pmatrix}
\qquad \Gamma^{\underline{t}}=
\begin{pmatrix}
i \sigma^1  &   0 \\
0 & i\sigma^1
\end{pmatrix}
\nonumber
\ee
\be
\Gamma^{\underline{x}}=
\begin{pmatrix}
-\sigma^2  &   0 \\
0 & \sigma^2
\end{pmatrix}
\qquad \Gamma^{\underline{y}}=
\begin{pmatrix}
0 &  \sigma^2 \\
\sigma^2 & 0
\end{pmatrix} .
 \nonumber
\ee

\noindent
Since in the gravity background the $\omega_{ab M}$ spin connection satisfies
\begin{equation} \frac{1}{4} \omega_{ab M} e^M_c \Gamma^c
 \Gamma^{ab} = \frac{1}{4} \Gamma^r \partial_r \ln \left( - g g^{rr} \right) ,
  \nonumber
\end{equation}
we can remove the spin connection from the
equations by introducing $ \binom{\psi_1}{\psi_2} :=  ( - g g^{rr} )^{1/4}  \spinor$.
The bulk Dirac equation decouples into two equations containing the two two-component spinors $\psi_1$ and $\psi_2$
{
%\footnotesize
\bea
\nonumber
 &&   \le[-\sqrt{g^{rr}} \sigma^3 \partial_r  - \mspinor + \eta \phi   \mp \sqrt{g^{xx}} i \sigma^2 k \ri. \hskip 2cm \\
  &  & \hskip 2cm   \le. + \sqrt{g^{tt}} \sigma^1 q A_t +\sqrt{g^{tt}} \omega \sigma^1 \ri]\psi_{1,2} = 0
\nonumber
\eea
}
%In the rest of the paper, we will focus on a single component and suppress the index on $\psi_i$.
At the UV $AdS_{4}$ boundary, any
solution to the equations of motion for the two-component spinor can be decomposed as follows,
\be
  \psi(r) \stackrel{r \rightarrow \infty}{\longrightarrow}
       \psi^+ r^{mR} \binom{ 0}{1}
    + \psi^- r^{-mR} \binom{ 1}{ 0} . \nonumber
\ee
The solution with {\it ingoing} boundary conditions at the horizon will be denoted $\psi_{\rm in}$. The {\it normalizable} and {\it non-normalizable} solutions will be denoted $\psi_\flat$ and $\psi_\sharp$, respectively. They satisfy
%$ \psi_\flat^+ = 0, \ \psi_\flat^- = 1, \ \psi_\sharp^+ = 1,  \ \psi_\sharp^- = 0$.
$ \psi_\flat^+ = 0$ and $\psi_\sharp^- = 0$.
Since the equation of motion is real, these solutions are also real.
The bulk Dirac equations for the two boundary spinor components $\psi_{1,2}$ have decoupled. Thus, the (diagonalized) retarded Green's function of the dual composite fermionic operators is given by \cite{Henningson:1998cd, Mueck:1998iz, Iqbal:2009fd, Faulkner:2009wj, Faulkner:2009am}
\be
  G_{\alpha\beta}(\omega, k) = %\gamma^{t} =
  \begin{pmatrix}
{\psi_{1,\rm in}^-(\omega, k) \ov \psi_{1,\rm in}^+(\omega, k)}  &   0 \\
0 & {\psi_{2,\rm in}^-(\omega, k) \ov \psi_{2,\rm in}^+(\omega, k)}
\end{pmatrix}
\ee
Here the label ``in'' indicates that the wavefunctions are ingoing.
 %and $\gamma^t = i\sigma^2$ is a boundary gamma matrix.
Note that in the boundary theory, the energy $\om$ is measured from the chemical potential.

We assume that there is a Fermi surface at a given momentum, i.e. at zero temperature the Green's function has a pole. This happens when the ``source'' component vanishes, $  \psi_{\rm in}^+(\omega=0, k_F) = 0$.
This equation defines the Fermi momentum $k_F$. As studied in detail in the papers \cite{Faulkner:2009wj, Faulkner:2009am}, the Green's function takes the following form,
\be
  G(\om, k) = {h_1 \ov k_\perp - {\om \ov v_F} + h_2 \mathcal{G}_{k_F}(\om)}
\ee
where $\mathcal{G}_{k_F}(\om)$ is the $AdS_2$ fermionic correlation function. In rescaled units, it is given by
\bea
  \mathcal{G}_{k}(\om) =  (4 \pi T)^{2 \nu} \frac{\( m - i k\) R_2+i q e_d + \nu}{ \( m - i k\) R_2+ i q e_d -  \nu} \qquad
   \\
   \nonumber
 \times
  {\Gamma (-2 \nu ) \Gamma (\half+\nu -\frac{i \omega }{2 \pi T }+i q e_d )\,  \Gamma \left(1+\nu -i
q e_d \right)\ov \Gamma (2 \nu )\Gamma \left(\frac{1}{2}-\nu -\frac{i \omega }{2 \pi T}+i q e_d
\right)\, \Gamma \left(1-\nu -i q e_d \right)}
\eea
where $\nu = {1\ov \sqrt{6}} \sqrt{m^2 + k^2 - {q^2 \ov 2}   } $.
At $T=0$, $\mathcal{G}_{k_F}(\om) \sim \om^{2\nu}$.

\subsection{ Bosonic Green's function  }

Generically, the low-energy behavior of the correlation function is dictated by the near-horizon $AdS_2$ geometry.  The scalar $AdS_2$ correlation function is \cite{Faulkner:2009wj},
\be
\chi(\om, T) = (4\pi T)^{2\nu_B} {\Gamma(-2\nu_B)\Gamma\le(\half+\nu_B-{i\om\over 2\pi T}\ri)\Gamma\le(\half+\nu_B\ri)  \over \Gamma(2\nu_B)\Gamma\le(\half-\nu_B-{i\om\over 2\pi T}\ri)\Gamma\le(\half-\nu_B\ri) } \nonumber
\ee
where $\nu_B = {1\ov \sqrt{6}} \sqrt{m^2 + {3\over 2}} $ in rescaled coordinates.

At low energies, the full $AdS_4$ correlation function can be constructed by a matching procedure similarly to the fermionic case. Fixing the charge, the Green's function will depend on the bulk mass of the field.

If the scalar mass is below the Breitenlohner-Freedman bound, then a bulk soliton interpolates between the minimum of the potential and the UV region where the scalar field is uncondensed. Deep in the infrared, a
new $\widetilde{AdS_2}$ describes the low-energy physics (see \cite{Iqbal:2010eh, Faulkner:2010gj}).

The paper \cite{Faulkner:2010gj} introduced double trace deformations in this scenario. In addition to the mass, double trace operators can be used to drive the system into a symmetry breaking phase.
The phase diagram is shown in FIG. \ref{fig:phase}. The two-dimensional parameter space is spanned by the scalar mass $\Delta m^2 = m^2 - m_c^2$ and the coefficient $\kappa$ of the double trace deformation.

In the following, we will briefly summarize the bosonic spectrum for the bifurcating and hybridized critical lines and for the marginal critical point.

\subsubsection{Marginal critical point}

If we set $m^2 = m_c^2$, then solving the Klein-Gordon equations in the IR $AdS_2$ region gives the IR correlator
\be
  \chi_{\rm IR}(\om, T)  = -\psi_0\left( \frac{1}{2}-\frac{i \omega }{2 \pi  T}\right) + \textrm{const.}
\ee
where $\psi_0$ is the digamma function. At the marginal critical point, this will be the full $AdS_4$ correlation function as well, see \cite{Iqbal:2011aj}.
The  function is analytic on the upper half plane, i.e. there is no condensation instability. The imaginary part is
\be
\Im \chi(\om, T)  =    {\pi  \over 2 } \tanh  { \omega \over 2T}
\ee
satisfies $\textrm{sign}(\om) \, \Im \chi(\om) \ge 0$.
In this approximation, the spectral function has no momentum dependence.
Note the two limits,
\be
\Im \chi =  \left\{ \begin{array}{ll}
 {\pi  \over 2 } \cdot { \omega \over T} & {\rm for} \quad {|\om|\over T} \rightarrow 0\\
 {\pi \over 2}  \cdot{\rm sign}\, \om & {\rm for} \quad {|\om|\over T} \rightarrow \infty
\end{array} \right.
\label{eqn:con}
\ee
Thus, in the low energy limit, the bosonic spectrum bears a strong resemblance to the one proposed by Varma et. al. \cite{varma} for describing the anomalous transport properties of cuprates. This observation was also made in \cite{Iqbal:2011aj, Jensen:2011af}. See also \cite{sachdevt} for similar results.

\subsubsection{Bifurcating critical point}

\begin{figure}[h] \begin{center}
\includegraphics[scale=0.52]{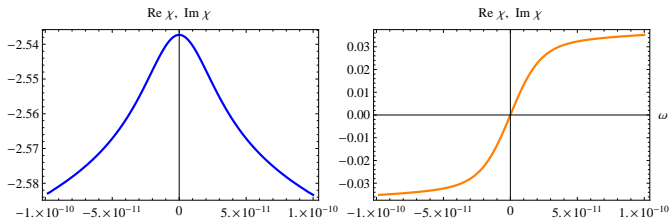}
\caption{\label{fig:bosoncorr} Real and imaginary parts of the boson response function $\chi_{\rm UV}$ at low energies ($\om = -{10 \, T} \ldots 10 \, T$; $T/\mu \sim 10^{-11}$).}
\end{center}
\end{figure}

At a bifurcating critical point, the full UV response function can be built by a matching procedure,
\be
  \chi(\om,T) \sim {1 \ov  c_1  - \log \om_c + \chi_{\rm IR}(\om,T) }
\ee
This form matches the numerically computed Green's function (see FIG. \ref{fig:bosoncorr}). For $\om \gg T$,  the imaginary part goes as
$ \Im \chi  \sim { \le( \log {\om } \ri)^{-2} }$.

\subsubsection{Hybridized critical point}

At the hybridized critical point, the $T=0$ correlation function takes the form \cite{Faulkner:2010gj}
\be
   \chi(\om, \vec k) \sim {1 \ov c_1 k^2 + c_2 \om^2 + c(k) \om^{2\nu_B} }
\ee
where $c_{1,2}$ are real and $c(k)$ is a complex function of the momentum. In the following, we will approximate it by a constant $c \equiv c(k=0)$. We will denote the real and imaginary part of this constant by $c'$ and $c''$, respectively.

\section{Fermion self-energy from bulk loop diagrams}

\begin{figure}[h] \begin{center}
\includegraphics[scale=0.56]{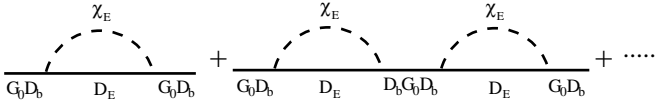}
\caption{\label{fig:sunsets} Bulk loop diagrams form a geometric series when the fermion momentum is at $k=k_F$. $D_E(r_1, r_2)$ and $\chi_E(r_1, r_2)$ are the fermion and boson bulk-to-bulk propagators, respectively. $G_0$ is the tree-level boundary fermionic Green's function. $D_\flat \equiv \spinor_\flat(r)$ denotes the normalizable fermionic wavefunction.}
\end{center}
\end{figure}

In this section, we compute the bulk self-energy diagrams in FIG. \ref{fig:sunsets}.
The bulk-to-bulk propagator for the boson ($\chi$) and for the fermion ($D$) are defined by
\bea
  \nonumber
  (\nabla^{(1)})^2\chi(r_1, r_2; \om, k) ={1\ov \sqrt{-g}} \delta(r_1-r_2) \\
  \nonumber
  (\Dslash^{(1)} -m) D(r_1, r_2; \om, k) ={1\ov \sqrt{-g}} \delta(r_1-r_2)
\eea
where the differential operators act on the $r_1$ radial coordinate.
A solution can be written down \cite{conduc} using normalizable ($\flat$) and and non-normalizable ($\sharp$) wavefunctions (henceforth, we are dropping insignificant constant factors)
\bwt
\bea
  \label{eqn:btb}
%  D(r_1, r_2; \om, k)  = {1\over W}   \spinor_{\flat}(r_>; \omega, k)   \spinor_{{\rm in}} (r_<; \omega, k)
  && D(r_1, r_2; \om, k)  = \left\{ \begin{array}{ll}
    \spinor_{\flat, \alpha}(r_1; \om, k) \le( G_{\alpha\beta}(\om,k) \overline{\spinor}_{\flat, \beta}(r_2; \om, k) - \gamma^t_{\alpha \beta} \overline{\spinor}_{\sharp,\beta}(r_2; \om, k) \ri) & \quad \textrm{if $r_1 > r_2$}\\
    \\
    \le( \spinor_{\flat, \alpha}(r_1; \om, k) G_{\alpha\beta}(\om,k)  -  {\spinor}_{\sharp,\alpha}(r_2; \om, k)  \gamma^t_{\alpha \beta} \ri) \overline{\spinor}_{\flat, \beta}(r_2; \om, k)& \quad \textrm{if $r_1 < r_2$}
  \end{array} \right.
\eea
\ewt
Here $\alpha, \beta = 1,2$ are boundary spinor indices, $\gamma^t$ is a boundary gamma  matrix and $G$ is the boundary retarded Green's function for the fermion. See \cite{conduc} for a derivation and more details on a similar calculation. The above solution parallels that of a scalar
\be
  \chi(r_1, r_2; \om, k)  = \phi_{\flat}(r_>; \omega, k)   \phi_{{\rm in}} (r_<; \omega, k)
\ee
where ``in'' indicates that the solution is infalling near the horizon.
The ingoing solution can be written as a linear combination of the non-normalizable and the normalizable wavefunctions
\be
  \phi_{{\rm in}}(r; \omega, k) =  \phi_{{\sharp}}(r; \omega, k) +  {\chi}(\omega, k)  \phi_{{\flat}}(r; \omega, k)
\ee
where $\chi$ is the boundary retarded Green's function for the scalar.
For real $\om$, the equations of motion are real. Since the $\sharp, \flat$ boundary conditions are also real, we can write
\be
  \Im \phi_{{\rm in}}(r; \omega, k) =   \phi_{{\flat}}(r; \omega, k) \, \Im  {\chi}(\omega, k)
\ee
and thus
\be
  \Im \chi(r_1, r_2; \om, k)  =  {  \phi_{\flat}(r_1; \omega, k)  \phi_{\flat}(r_2; \omega, k)  }   \Im  {\chi}(\omega, k)
  \nonumber
\ee
Note that this is an exact expression. Similarly, for the fermion we have
  \cite{conduc, Faulkner:2010zz}
\be
    \label{eqn:factor}
    \rho(r_1, r_2; \om, k)  =  {  \spinor_{\flat, \alpha}(r_1; \omega, k)   \rho_{\alpha \beta}(\omega, k) \overline\spinor_{\flat, \beta}(r_2; \omega, k)  }
\ee
where the bulk ($\rho$) and boundary ($\rho_{\alpha \beta}$) spectral functions  are defined to be the difference between the corresponding retarded and advanced propagators: $\rho = -i(D_R -D_A)$ and $\rho_{\alpha \beta} = -i(G_R- G_A)$.

We now want to compute the sum of diagrams in FIG. \ref{fig:sunsets}.
In field theory, the diagrams form a geometric series and the result can be simply obtained from the one-loop diagram. Here we are dealing with bulk diagrams which do not necessarily form a geometric series. The problem arises because propagators between loops do not generically factorize: $D(r_1, r_2) \ne X(r_1) Y(r_2)$.

However, at small temperatures the bulk-to-bulk propagator (\ref{eqn:btb}) {\it does} factorize on the Fermi surface,
\be
  \label{eqn:factorize}
  D(r_1, r_2; \om , k) \approx  \spinor_{\flat, \alpha}(r_1; \om, k)   G_{\alpha\beta}(\om,k) \overline{\spinor}_{\flat, \beta}(r_2; \om, k)
\ee
with  $\om \approx 0$ and $ k \approx k_F$. This occurs because $G$ blows up and we can neglect the non-normalizable piece in (\ref{eqn:btb}).

The factorization property of bulk-to-bulk propagators implies that the diagrams form a geometric series,
\be
  \overline{G} =   {G} +
    {G} \Sigma  {G}+
    {G} \Sigma {G}  \Sigma {G} + \ldots
\ee
where $\overline{G}$ is the resummed boundary Green's function. The above form of the series can be understood as follows. The first term is just the tree-level diagram. The second term is the one-loop diagram. In Euclidean signature in 3+1 bulk dimensions,
\bwt
{%\footnotesize
\bea
  \nonumber
  &&  G_{\alpha\kappa_1}(i\Om, \vec k ) \Sigma_{\kappa_1\kappa_2}(i\Om, \vec k ) G_{\kappa_2\beta}(i\Om, \vec k ) =  \\
  && =  T \sum_{i\Om'_m}
  \int  {d^2 q \over (2\pi)^2}    dr_1 \sqrt{g(r_1)}   dr_2 \sqrt{g(r_2)} \,
  \tr\le( D_{\rm E}(r_1, r_2;  i\Om - i\Om'_m  ,\vec{k}-\vec{q}) D_{\rm E, \beta}(r_2; i\Om, \vec k )
   \chi_{\rm E}(r_2, r_1; i\Om'_m, |q|)  \overline D_{\rm E, \alpha}(r_1; i\Om, \vec k )
   \ri)
   \nonumber
\eea
}
where $D_E(r_1, r_2)$ and $\chi_E(r_1, r_2)$ are the fermionic and bosonic Euclidean bulk-to-bulk propagators, respectively. $D_{\rm E, \alpha}(r)$ and $\overline D_{\rm E, \alpha}(r)$ are the bulk-to-boundary and boundary-to-bulk propagators. Finally, ``$\tr$'' denotes the trace in bulk spinor indices, and $\Om'_m = 2\pi m T$ with $m\in \ZZ$.
At low energies and near the Fermi surface, the bulk-to-boundary propagators can be expressed as
\be
  D_{\rm E, \alpha}(r;  i\om, \vec k) \approx   \spinor_{\flat,\beta}(r;  i\om, \vec k) G_{\beta\alpha}(i\om, \vec k)\qquad \overline D_{\rm E, \alpha}(r;  i\om, \vec k) \approx
  G_{\alpha\beta}(i\om, \vec k) \overline\spinor_{\flat,\beta}(r;  i\om, \vec k)
\ee
Using this we have
{%\footnotesize
\bea
  \nonumber
  && (G  \Sigma  G)_{\alpha\beta}  \approx
    T \sum_{i\Om'_m}
  \int  {d^2 q \over (2\pi)^2}    dr_1 \sqrt{g(r_1)}   dr_2 \sqrt{g(r_2)} \times\\
  &&\times\tr\le( D_{\rm E}(r_1, r_2;  i\Om - i\Om'_m  ,\vec{k}-\vec{q})
   \spinor_{\flat,\kappa_2}(r_2;  i\Om, \vec k)  G_{\kappa_2\beta}(i\Om, \vec k)
   \chi_{\rm E}(r_2, r_1; i\Om'_m, |q|)
    G_{\alpha \kappa_1}(i\Om, \vec k)
   \overline\spinor_{\flat,\kappa_1}(r_1;  i\Om, \vec k)
   \ri)
     \nonumber
%   T \sum_{i\Om'_m}
%  \int  {d^{d-1} q \over (2\pi)^2}  \int dr_1 \sqrt{g_1}   dr_2 \sqrt{g_2} \, D_{\rm E}(r_1, r_2;  i\Om - %i\Om'_m  ,\vec{k}-\vec{q})
%   \chi_{\rm E}(r_1, r_2; i\Om'_m, |q|) \spinor_{\flat}(r_1;  i\Om, \vec k) G(i\Om, \vec k )  %\spinor_{\flat}(r_2;  i\Om, \vec k) G(i\Om, \vec k )
\eea
}
Higher loop Feynman diagrams also contain bulk-to-bulk propagators between the loops. On the Fermi surface, they factorize into two normalizable wavefunctions, see (\ref{eqn:factorize}), and the sum of diagrams forms a geometric series.

In order to sum the series, we compute the  ``half-amputated'' self-energy diagram. This does not contain the two $G_{\alpha\beta}(i\Om, \vec k )$ functions from the bulk-to-boundary propagators, but does include the
corresponding $\spinor_{\flat,\alpha}(r_{1,2})$
normalizable spinor wavefunctions.
{%\footnotesize
\be
  \nonumber
  \Sigma_{\alpha\beta}  =
    T \sum_{i\Om'_m}
  \int  {d^2 q \over (2\pi)^2}    dr_1 \sqrt{g(r_1)}   dr_2 \sqrt{g(r_2)}
 \tr\le( D_{\rm E}(r_1, r_2;  i\Om - i\Om'_m  ,\vec{k}-\vec{q}) \spinor_{\flat,\beta}(r_2;  i\Om, \vec k)
   \chi_{\rm E}(r_2, r_1; i\Om'_m, |q|) \overline\spinor_{\flat,\alpha}(r_1;  i\Om, \vec k)
   \ri)
     \nonumber
%   T \sum_{i\Om'_m}
%  \int  {d^{d-1} q \over (2\pi)^2}  \int dr_1 \sqrt{g_1}   dr_2 \sqrt{g_2} \, D_{\rm E}(r_1, r_2;  i\Om - %i\Om'_m  ,\vec{k}-\vec{q})
%   \chi_{\rm E}(r_1, r_2; i\Om'_m, |q|) \spinor_{\flat}(r_1;  i\Om, \vec k) G(i\Om, \vec k )  %\spinor_{\flat}(r_2;  i\Om, \vec k) G(i\Om, \vec k )
\ee
}
Using the spectral decomposition of the bulk-to-bulk propagators, we rewrite $\Sigma$
{\footnotesize
\be
\nonumber
   \Sigma_{\alpha\beta}(i\Om, \vec{k} ) =  \int_{-\infty}^{\infty} {d\om_1 \over 2\pi}  {d\om_2 \over 2\pi} T \sum_{i\Om'_m}
  \int  {d^2 q \over (2\pi)^2} dr_1 \sqrt{g_1}    dr_2 \sqrt{g_2}
   \tr\le( {\rho(r_1, r_2; \om_1  , \vec{k}-\vec{q})  \over i\Om - i\Om'_m - \om_1}
   \spinor_{\flat,\beta}(r_2;  i\Om, \vec k)
{   \Im \chi(r_1, r_2; \om_2, |q|)
   \over   i\Om'_m-\om_2  }  \overline\spinor_{\flat,\alpha}(r_1;  i\Om, \vec k)
   \ri)
\ee
}
From eqn. (\ref{eqn:factor}), the bulk-to-bulk spectral function $\rho$ can be expressed using the one in the boundary theory,
{\footnotesize
\bea
   && \Sigma_{\alpha\beta}(i\Om, \vec{k} ) =  \int_{-\infty}^{\infty} {d\om_1 \over 2\pi}  {d\om_2 \over 2\pi} T \sum_{i\Om'_m}
  \int  {d^2 q \over (2\pi)^2} dr_1 \sqrt{g_1}    dr_2 \sqrt{g_2} \times \\
  &&
   \tr\le( { \spinor_{\flat,\sigma_1}(r_1; \om_1  , \vec{k}-\vec{q}) \rho_{\sigma_1 \sigma_2}(\om_1  , \vec{k}-\vec{q})
    \overline \spinor_{\flat,\sigma_2}(r_2; \om_1  , \vec{k}-\vec{q}) \over i\Om - i\Om'_m - \om_1}
   \spinor_{\flat,\beta}(r_2;  i\Om, \vec k)
{   \phi_{\flat}(r_1;  \om_2  ,|q|)
   \phi_{\flat}(r_2;  \om_2  ,|q|) \Im \chi(  \om_2, |q|)
   \over   i\Om'_m-\om_2  }  \overline\spinor_{\flat,\alpha}(r_1;  i\Om, \vec k)
   \ri)
\nonumber
\eea
}
%\clearpage
The integrand has factorized into $r_1$ and $r_2$ dependent parts.
Evaluating the sum over $\Om'_m$ gives
{%\footnotesize
\bea
    \label{eqn:beforevertex}
&&  \hskip -0.0cm \Sigma_{\alpha\beta}(i\Om, \vec{k} ) =  \int_{-\infty}^{\infty} {d\om_1 \over 2\pi} {d\om_2 \over 2\pi}   \int  {d^2 q \over (2\pi)^2}  \int dr_1 \sqrt{g_1} dr_2 \sqrt{g_2}
{ \tanh\le({\om_1 \over 2T}\ri)+ \coth\le({\om_2 \over 2T}\ri) \over \Om - \om_1 - \om_2 + i \epsilon}
 \phi_{\flat}(r_1;  \om_2  ,|q|)
   \phi_{\flat}(r_2;  \om_2  ,|q|)
\times
    \nonumber
    \\
&&  \times \Im \chi(  \om_2, |q|) \,
 \tr \le[
 \spinor_{\flat,\sigma_1}(r_1; \om_1  , \vec{k}-\vec{q})
  { \rho_{\sigma_1 \sigma_2}(\om_1  , \vec{k}-\vec{q})
    \overline \spinor_{\flat,\sigma_2}(r_2; \om_1  , \vec{k}-\vec{q}) }
   \spinor_{\flat,\beta}(r_2;  i\Om, \vec k)
 \overline\spinor_{\flat,\alpha}(r_1;  i\Om, \vec k)
 \ri]
 \eea
}
Let us now pick a basis where $\rho_{\alpha\beta}(\om, \vec k)$ is diagonal and fix the two basis vectors $\spinor_{\flat,\alpha=1,2}(r; \om, \vec k)$ for the $\vec k$ direction. For other momentum vectors in the $x-y$ plane we define a basis for the normalizable wavefunctions
\be
   \spinor_{\flat,\alpha}\le(r; \om, \mathcal{R(\varphi)}\vec k \ri) :=   e^{i\varphi \ov 2} R_{\alpha\beta}(\varphi)
    \spinor_{\flat,\beta}(r; \om, \vec k)
\ee
where $\mathcal{R}$ and $R$ are the boundary vector and spinor rotations, respectively. Note that we have included a phase in order to cancel the monodromy $R(2\pi) = -{\rm Id}$.

Since $\vec{k}-\vec{q}$ is not parallel to $\vec{k}$, the spectral density $\rho_{\alpha\beta}(\vec{k}-\vec{q})$ in (\ref{eqn:beforevertex}) has to be diagonalized,
\be
 \nonumber
  \spinor_{\flat,\alpha}(r_1; \om_1  , \vec{k}-\vec{q})\rho_{\alpha\beta}(\om_1  , \vec{k}-\vec{q}) \overline \spinor_{\flat,\beta}(r_2; \om_1  , \vec{k}-\vec{q})
   =  \le( e^{i\varphi \ov 2} R_{\alpha\sigma_1} \spinor_{\flat,\sigma_1}  \ri)
    \underbrace{
     % e^{i\varphi \ov 2}
      R_{\alpha\sigma_2}
    \rho_{\sigma_2 \sigma_3}(\vec{k}-\vec{q})
    %e^{-i\varphi \ov 2}
    \le(R^{-1}\ri)_{\sigma_3\beta}
    }_{\rho_{\alpha\beta}(\hat k |\vec{k}-\vec{q}|)}
    \overline{ e^{i\varphi \ov 2} R_{\beta\sigma_4} \spinor_{\flat,\sigma_4} }
%     \le( e^{i\varphi \ov 2} R_{\alpha\sigma_1} \spinor_{\flat,\sigma_1}(r_1; \om_1  , \vec{k}-\vec{q}) \ri)
%    \le( e^{i\varphi \ov 2} R_{\alpha\sigma_2}
%    \rho_{\sigma_2 \sigma_3}(\om_1  , \vec{k}-\vec{q})
%    e^{-i\varphi \ov 2} R^{-1}_{\sigma_3\beta}   \ri)
%    \overline{ e^{i\varphi \ov 2} R_{\beta\sigma_4} \spinor_{\flat,\sigma_4} }(r_2; \om_1  , \vec{k}-\vec{q})
\ee
where $ \hat k = {\vec k / |\vec k|}$, and $\varphi$ is the angle between $\vec{k}-\vec{q}$ and $\vec{k}$. The rotated wavefunctions can be expressed using the original ones in the $\vec k$ direction
\be
   e^{i\varphi \ov 2} R_{\alpha\beta} \spinor_{\flat,\beta}\le(r_1; \om, \vec{k}-\vec{q}\ri)  =
    \spinor_{\flat,\alpha}\le(r_1; \om, \hat k |\vec{k}-\vec{q}|\ri) % \qquad
\ee
Then, we can rewrite (\ref{eqn:beforevertex}) as
{%\footnotesize
\bea
    \label{eqn:beforevertex2}
&&  \hskip -0.0cm \Sigma_{\alpha\beta}(i\Om, \vec{k} ) =  \delta_{\alpha\beta} \int_{-\infty}^{\infty} {d\om_1 \over 2\pi} {d\om_2 \over 2\pi}   \int  {d^2 q \over (2\pi)^2}  \int dr_1 \sqrt{g_1} dr_2 \sqrt{g_2}
{ \tanh\le({\om_1 \over 2T}\ri)+ \coth\le({\om_2 \over 2T}\ri) \over \Om - \om_1 - \om_2 + i \epsilon}
 \phi_{\flat}(r_1;  \om_2  ,|q|)
   \phi_{\flat}(r_2;  \om_2  ,|q|)
\times
        \nonumber
    \\
&&  \times \Im \chi(  \om_2, |q|) \,
 \tr \le[
 \spinor_{\flat,\alpha}(r_1; \om_1  ,\hat k |\vec{k}-\vec{q}|)
  { \rho_{\alpha\alpha}(\om_1  , \hat k |\vec{k}-\vec{q}|)
    \overline \spinor_{\flat,\alpha}(r_2; \om_1  , \hat k |\vec{k}-\vec{q}|) }
   \spinor_{\flat,\alpha}(r_2;  i\Om, \vec k)
 \overline\spinor_{\flat,\alpha}(r_1;  i\Om, \vec k)
 \ri]
 \eea
}
There is no summing over $\alpha$. The momentum arguments inside the trace are now all parallel vectors.

\subsection{The effective vertex}

We will be concerned with the boundary component $\alpha$ for which there is a Fermi surface, {\it i.e.} $G^{-1}_{\alpha\alpha}(0,k_F) = 0$ at $T=0$.
The corresponding self-energy in (\ref{eqn:beforevertex2}) can be rewritten (we will suppress $\hat k$ in the following)
\bea
 \label{eqn:master}
  &&  \Sigma_{\alpha\alpha}(\Om, \vec{k} ) = \sum_{\alpha=1,2} \int_{-\infty}^{\infty} {d\om_1 \over 2\pi} {d\om_2 \over 2\pi}
  {d^{d-1} q \over (2\pi)^2} \,
   \Lambda_{\flat, \alpha}\le(\Om, |\vec k|; \,  \om_1, |\vec k - \vec q|;  \, \om_2, |\vec q| \ri)
   \,   \Lambda_{\flat, \alpha}\le( \om_1, |\vec k - \vec q|;  \, \Om, |\vec k|;  \, \om_2, |\vec q| \ri)  \times
      \nonumber
   \\
   && \hskip 3cm \times \rho_{\alpha\alpha}( \om_1, |\vec{k}-\vec{q}|) \Im \chi(  \om_2, |q|) { \tanh\le({\om_1 \over 2T}\ri)+ \coth\le({\om_2 \over 2T}\ri) \over \Om - \om_1 - \om_2 + i \epsilon}
\eea
where we defined the effective vertex using normalizable two-component wavefunctions
\be
 \nonumber
  \Lambda_{\flat, \alpha}(\om_1, \vec k_1; \, \om_2, \vec k_2; \,  \om_3, \vec k_3) = \int dr \sqrt{g} \, \overline\psi_{\flat,\alpha}(r; \om_1, \vec k_1) \psi_{\flat,\alpha}(r; \om_2, \vec k_2)  \phi_{\flat}(r; \om_3, \vec k_3)
\ee
%and $\overline\psi = i \psi^\dagger \sigma^1$.
We will use the $2\times 2$ gamma matrices $\gamma^{\underline{t}} = i \sigma^1, \gamma^{\underline{r}} =\sigma^3, \gamma^{\underline{x}} = -\sigma^2$.
In order to compute $\Im \Sigma$, we do the $\om_2$ integral
{
\bea
 \nonumber
 && \Im \, \Sigma(\Om, \vec{k} ) = \int_{-\infty}^{\infty} {d\om_1 \over 2\pi}{d^{d-1} q \over (2\pi)^2}
 \,
  \Lambda_{\flat,\alpha}\le(\Om, |\vec k|; \, \om_1, |\vec k - \vec q|; \,   \Om - \om_1, |\vec q|\ri)
 \Lambda_{\flat,\alpha}\le( \om_1, |\vec k - \vec q|; \, \Om, |\vec k|; \,  \Om - \om_1, |\vec q|\ri) \times
 \\
&& \hskip 3cm \times  \rho_{\alpha\alpha}( \om_1, |\vec{k}-\vec{q}|) \Im \chi(  \Om - \om_1 , |q|) \left[ \tanh\le({\om_1 \over 2T}\ri)+ \coth\le({\Om - \om_1  \over 2T}\ri) \right]
  \label{eqn:master2}
\eea
}

\clearpage

\ewt

At low temperatures, the hyperbolic functions suppress the integrand outside the window $\om_1 = [0\ldots \Om]$. The fermion spectral function can be approximated by $\Im G_R \approx Z \delta(k_\perp - \Re \Sigma_0)$. This effectively restricts the momentum integral to the vicinity of the Fermi surface and we have $|\vec k| \approx k_F, |\vec k - \vec q| \approx k_F$.

\subsubsection{Infrared divergences: $T \ll  \zeta^{-1}  \ll \om \ll \mu$}

A potential divergence may come from the deep IR region: $T \ll \zeta^{-1} \ll \om$.
(See \cite{Hung:2010pe} for divergences in a similar effective vertex in $AdS_n$).
Let us fix $T=0$ and use the $AdS_2$ metric: $ds^2 = {-d\tau^2 +d\zeta^2 \ov \zeta^2}$. The neutral boson and charged fermion wavefunctions can be expressed using Hankel and Whittaker functions, respectively \cite{Faulkner:2011tm}
\be
   \phi(\zeta) =  c_{{\rm out}}{\rm H}^{(1)}_{\nu}(\omega \zeta )
   +  c_{{\rm in}}{\rm H}^{(2)}_{\nu}(\omega \zeta )
\ee
 \bea
  \nonumber
&& \hskip -0.5cm \tilde \spinors(\zeta)  \equiv
 {1\over \sqrt 2} (1 + i  \sigma^1)
%\left( - g g^{\zeta\zeta}\right)^{-1/4}
\spinors
= % \zeta^{-1/2} \left[
 c_{{\rm out}}
 {\rm W}_{-{\sigma^3\over 2} - i q, \nu}(2 i \omega \zeta)
 \binom{  \tilde m - i m }{ -1}
 \\
 && \hskip 3.2cm + c_{{\rm in}} {\rm W}_{{ \sigma^3\over 2} + i q, \nu}(-2 i \omega \zeta)
 \binom{ - 1 }{ \tilde m + i m} %\hskip 1cm
%\right]
\nonumber
\eea
where the notation $\sigma^3$ in the index of the Whittaker function
indicates $\pm 1 $ when acting on the top/bottom component of the spinor, and in our units $\tilde m = -(-1)^{\alpha} k$.

Let us first look at the effective vertex with ingoing/outgoing boundary conditions for the wavefunctions.
For $\om \zeta \gg 1$, the ingoing wavefunctions
\be
  \tilde\spinors_{\rm in}^{\pm}(\zeta ; \om) \sim \zeta^{\pm \half} e^{i\om \zeta} \zeta^{iq}
    \qquad \phi_{\rm in}(\zeta; \om)\sim  \zeta^{0} e^{i\om \zeta}
  \nonumber
\ee
where the two signs correspond to the two spinor components.
The outgoing wavefunctions satisfy $\spinors_{\rm out} = \le(\spinors_{\rm in}\ri)^*$, $\phi_{\rm out} = \le(\phi_{\rm in}\ri)^*$,
\be
  \tilde\spinors_{\rm out}^{\pm}(\zeta ; \om) \sim \zeta^{\mp \half} e^{-i\om \zeta} \zeta^{-iq}
    \qquad \phi_{\rm out}(\zeta; \om)\sim  \zeta^{0} e^{-i\om \zeta}
  \nonumber
\ee

The Yukawa coupling can be written as
\be
%  \overline\spinors \spinors \phi = \half \spinors^\dagger (1 -  \Gamma^{\underline{t}})  \Gamma^{\underline{t}} %(1+ \Gamma^{\underline{t}}) \spinors \phi = \tilde \spinors^\dagger \Gamma^{\underline{t}} \tilde\spinors \phi
  \overline\spinors \spinors \phi = \spinors^\dagger i\sigma^1\spinors \phi =  \spinors^\dagger {1 -  i\sigma^1 \ov \sqrt{2}}  i\sigma^1 {1 +  i\sigma^1 \ov \sqrt{2}} \spinors \phi = i\tilde \spinors^\dagger \sigma^1 \tilde\spinors \phi
  \nonumber
\ee
%where we used four-component spinors.
From the last form we see that the Yukawa coupling combines $\tilde\spinors^{+}$ with $\tilde\spinors^{-}$. For ingoing wavefunctions this gives
\be
\nonumber
 \overline\spinors_{\rm in} \spinors_{\rm in} \phi_{\rm in} \sim
 i \le(\tilde\spinors_{\rm in}^{+}\ri)^*  \tilde\spinors_{\rm in}^{-} \phi_{\rm in} +
 i \le(\tilde\spinors_{\rm in}^{-}\ri)^*  \tilde\spinors_{\rm in}^{+} \phi_{\rm in}
 \sim\zeta^0
\ee
Note that the $\zeta^{\pm iq}$ factors have canceled each other.
There is no divergent contribution to the effective vertex
\be
 \nonumber
   \Lambda^{IR}_{\rm in}(\Om, \om_1, \om_2 , k_F)  = %\hskip 0.6cm
   %\hskip 0.6cm
   \int_{cutoff}^\infty { d\zeta \over \zeta^2}
  \overline\spinors_{\rm in} \spinors_{\rm in} \phi_{\rm in}  < \infty
  %\le(\zeta^{+\half} e^{i\Om \zeta}\ri) \le(\zeta^{-\half} e^{i\om_1 \zeta}\ri) \le(\zeta^{0} e^{i\om_2 %\zeta}\ri) < \infty
   \nonumber
\ee
Similar results apply to outgoing wavefunctions.
Let us now turn to normalizable wavefunctions
\be
  \spinors_{\flat} = { \spinors_{\rm in} - \spinors_{\rm out} \ov 2i \Im G}
\ee
The Yukawa coupling contains various terms
\be
\nonumber
 \overline\spinors_{\flat} \spinors_{\flat} \phi  \sim
  \overline\spinors_{\rm in} \spinors_{\rm in} \phi   +  \overline\spinors_{\rm out} \spinors_{\rm out} \phi -\overline\spinors_{\rm in} \spinors_{\rm out} \phi - \overline\spinors_{\rm out} \spinors_{\rm in} \phi
\ee
The last two terms couple the spinor components $\tilde\spinors_{\rm in}^{+}$ to $\tilde\spinors_{\rm out}^{-}$, which scales for large $\zeta$ as
\be
  \tilde\spinors_{\rm in}^{+} \tilde\spinors_{\rm out}^{-} \sim \zeta^{1} e^{i \om \zeta} \zeta^{2i q}
\ee
As $\zeta \to \infty$, this is more divergent than the purely ingoing case.
The effective vertex has a logarithmic divergence at $q=0$. For positive fermion charges,
\be
 \nonumber
   \Lambda^{IR}_{\flat}  =
   \int { d\zeta \over \zeta^2}  \overline\spinors_{\flat} \spinors_{\flat} \phi_{\flat}
     \sim \int { d\zeta \over \zeta}   e^{i (0+i\varepsilon) \zeta} \zeta^{2i q} < \infty
\ee
Note that the finiteness of the vertex crucially depends on the spinor index structure and is not guaranteed for other types of interactions.

The UV region ({\it i.e.} the bulk of $AdS_4-RN$) contributes a constant to the effective vertex.
We have seen that the contribution from the region $\zeta^{-1} \ll \om$ is not divergent.
It remains to study potential divergences from the region $\om \ll \zeta^{-1}$.

\subsubsection{Infrared divergences: $ T, \om \ll \zeta^{-1} \ll \mu $}

In the bulk, $\psi_{\flat}$ and $\phi_{\flat}$ are temperature-independent to first order. These functions match onto those in the near-horizon $AdS_2$ region:
\be
  \spinors_{\flat}^{\pm}(\zeta ; 0, k_F) \sim  \le({\zeta \over T}\ri)^{\half-\nu_F} \qquad \phi_{\flat}(\zeta; 0,0)\sim \le({\zeta \over T}\ri)^{\half-\nu_B}
  \nonumber
\ee
where $\zeta \equiv {T \ov r-1}$ is the rescaled near-horizon coordinate.
The IR contribution then scales as
\bea
 \nonumber
  && \Lambda_{IR}(0,0,k_F)  = \int_{cutoff}^{horizon} dr \sqrt{-g} \, \overline\spinors_{\flat}(r) \spinors_{\flat}(r) \phi_\flat(r)   \\
  &&  \sim \int_{T \ov T_{c}}^\infty d\zeta{ T \over \zeta^2}  \le({\zeta \over T}\ri)^{1-2\nu_F}   \le({\zeta \over T}\ri)^{\half-\nu_B}
  \sim  T^{2\nu_F-\half+\nu_B} %\int  {d\zeta \over \zeta}  \zeta^{\half-2\nu}
  \nonumber
\eea
Here $T_c$ is a constant which sets the cutoff that separates the IR and UV regions.
Thus, the contribution from the IR $AdS_2$ is not singular in the zero temperature limit if $\nu_F >  {1 \ov 4}-{\nu_B \ov 2}$.

Note that when the internal fermion propagator is not on the Fermi surface, the corresponding wavefunction $\spinors_{\flat}^{\pm}(\zeta ; 0, k)$ has a piece that goes like $ \le({\zeta \over T}\ri)^{\half+\nu_F}$. The effective vertex then scales as $T^{\nu_B-\half}$ which is divergent when $\nu_B < \half$. However, the full expression for $\Im \Sigma$ contains a factor $\Im G_R \sim T^{2\nu_F}$ which -- for large enough $\nu_F$ -- renders this contribution harmless.

Finally, we can substitute
\be
  \Lambda(\Om, \om_1, \Om - \om_1 , k_F) \rightarrow  \Lambda(0, 0,0, k_F) \equiv \Lambda(k_F)
\ee
The resulting expression is now familiar from finite temperature quantum field theories
\bea
 \nonumber
  \Im \, \Sigma(\Om, \vec{k} ) \approx  \Lambda(k_F)^2 \int_{-\infty}^{\infty} {d\om_1 \over 2\pi}
    {d^{d-1} q \over (2\pi)^2}
  \Im {G}_{\rm R}( \om_1, |\vec{k}-\vec{q}|) \\
  \times\Im \chi(  \Om - \om_1 , |q|) \left[ \tanh\le({\om_1 \over 2T}\ri)+ \coth\le({\Om - \om_1  \over 2T}\ri) \right] \hskip 1cm
  \label{eqn:master3}
\eea
The only consequence of the radial direction is that we had to replace the interaction by the effective vertex $\Lambda$.

\section{Fermion self-energy results}

In this section we will use the previous results to compute the one-loop bosonic contribution to the fermion self-energy in various regions of the phase diagram FIG. \ref{fig:phase}. As it is clear from eqn. (\ref{eqn:master2}), this will be an ordinary field theory calculation where the interaction vertices are determined by radial integrals in the bulk.

\subsection{Marginal critical point}

The boson spectral function assumes the momentum-independent scale-invariant form,
\be
  \label{eqn:tanh}
   \Im \chi(  \om, \vec k, T) = \tanh\le( {\om \ov 2T } \ri)
\ee
This approximation is valid for small momenta.
%The $\om_{1,2}$ and $q$ integrals can be done with usual techniques.
Near the Fermi surface the fermionic spectral function becomes a Dirac delta,
\be
  \Im G_{\rm R, 0}(k_F + k_\perp, \om) \approx Z \delta(k_\perp - \Re \Sigma_0(\om))
\ee
It is easy to evaluate the momentum integral, since only $G_{\rm R, 0}$ contains $q$. We get,
{\footnotesize
\be
  \Im  \Sigma(\Om) \sim \int_{-\infty}^{\infty} {d\om_1 \over 2\pi}
     \tanh {\Om - \om_1 \ov 2T }   \le( \tanh{\om_1 \over 2T}+ \coth{\Om - \om_1 \over 2T}  \ri)
\ee
}
After performing the $\om_1$ integral we get %can be performed,
\be
   \Im \Sigma(\Om) \sim \Om \coth  {\Om \over 2T }
\ee

The real part can be computed using the Kramers-Kronig relation. The full result is,
\be
  \Sigma(\om, T) = \om \le[ {i \pi T \ov \om } +\log {\om_c \ov 2\pi T} -\psi_0\le(1-{i \om \ov 2 \pi T}\ri)\ri]
\ee
where $\psi_0$ denotes the digamma function. The term involving the UV scale $\om_c$ renders the large frequency limit well-defined, {\it i.e.} temperature independent.
This function is a smooth version of the original marginal Fermi liquid self-energy:
\be
  \Sigma_{\rm MFL}(\om, T) \sim \om \log {x \over \omega_c} - i {\pi \over 2} x
\ee
where $x \equiv \max(\om, T)$.

\subsection{Bifurcating critical point}

Let us consider the $T \ll \om$ limit. The bosonic spectral function is \cite{Iqbal:2011aj}
\be
   \Im \chi(  \om) \sim {{\pi \ov 2} \log {\om_a \ov \om_b}  \ov \log^2 {\om \ov \om_a} + {\pi^2 \ov 4} }
\ee
This is again independent of the momentum.
At $T=0$, the factor $\tanh{\om_1 \over 2T}+ \coth{\Om - \om_1 \over 2T}$   has support in the interval $\om_1 \in (0 \ldots \Om)$. Thus,
\be
  \Im  \Sigma(\Om) \sim \int_{0}^{\Om} {d\om_1 \over 2\pi}
      {1 \ov \log^2 {\Om - \om_1 \ov \om_a} + {\pi^2 \ov 4} }
\ee
At large $\om_a$ and  $\Om \ll \om_a$  we get,
\be
  \label{eqn:bifurself}
  \Im  \Sigma(\Om) \sim  {\Om \ov \log^2 {\Om \ov \om_a}  }
\ee

\subsection{Hybridized critical point in 2d }

In this case, we have to take into account the momentum dependence of the susceptibility. The fermionic self-energy,
\bea
  &&  \Sigma(\Om, \vec{k} ) \approx  \Lambda(k_F)^2\int_{-\infty}^{\infty} {d\om_1 \over 2\pi} {d\om_2 \over 2\pi}
  \int  {d^2 q \over (2\pi)^2} \\
  &&
     \hskip -0.5cm  \times \Im {G}_{\rm R, 0}( \om_1, |\vec{k}-\vec{q}|) \Im \chi(  \om_2, |q|) { \tanh\le({\om_1 \over 2T}\ri)+ \coth\le({\om_2 \over 2T}\ri) \over \Om - \om_1 - \om_2 + i \epsilon}
      \nonumber
\eea

At $T=0$, the imaginary part becomes
\bea
   && \hskip -0.5cm \Im \Sigma(\Om, \vec{k} ) \sim  {\rm sgn \, \Om } \, \Lambda(k_F)^2  \int_{0}^{\Om} {d\om_1 \over 2\pi}
  \int  {d^2 q \over (2\pi)^2} \\
  &&
      \times \Im {G}_{\rm R, 0}( \om_1, |\vec{k}-\vec{q}|) \Im \chi(  \Om-\om_1, |q|)
      \nonumber
\eea

\begin{figure}[h] \begin{center}
\includegraphics[scale=1]{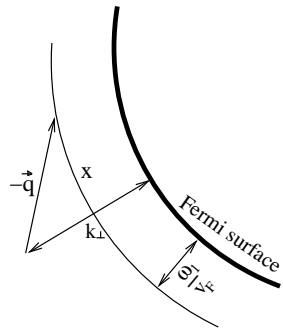}
\caption{\label{fig:mom} Variables in momentum space.}
\end{center}
\end{figure}

The Dirac delta in $\Im G$ restricts the momentum integration such that $|\vec{k}-\vec{q}| -k_F= {\om_1 \over  v_F}$. Let us ignore the curvature of the Fermi surface and denote the tangential component of the momentum by $x$ (see FIG. \ref{fig:mom}).
 Then,
{
\bea
  &&  \Im  \Sigma(k_\perp, \Om) \sim  {\rm sgn \, \Om } \, \Lambda(k_F)^2 \\
  \nonumber
  &&  \hskip -0.3cm  \times \int_{0}^{\Om} {d\om_1 \over 2\pi} \int_{-\infty}^{\infty} dx \,
    \le. \Im \chi(\om, q) \ri.\bigg|_{\om = \Om-\om_1; \ q = \sqrt{\le(k_\perp - {\om_1 \ov v_F} \ri)^2 + x^2}}
\eea
}

We will need the imaginary part of the susceptibility,
\be
   \Im \chi(\om, \vec k) \sim {  c'' \om^{2\nu_B}  \ov \le(c_1 k^2  + c_2 \om^2+   c' \om^{2\nu_B} \ri)^2  + c''^2\om^{4\nu_B}  }
\ee
In the following, we will assume $\Om > 0$.

\begin{figure}[h] \begin{center}
\includegraphics[scale=0.8]{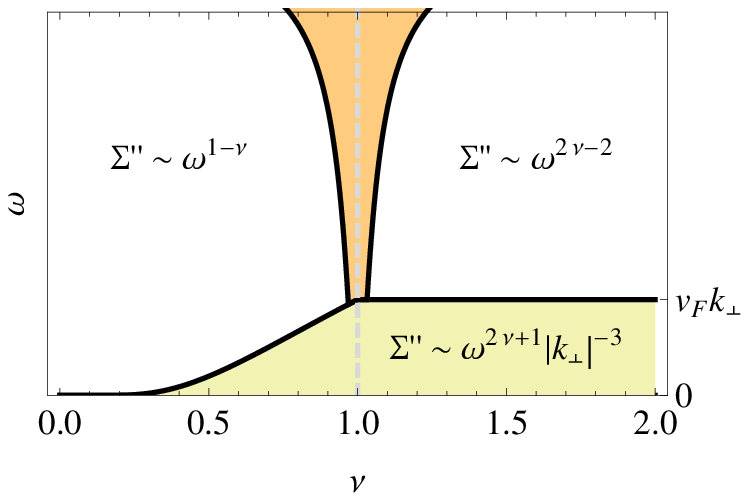}
\caption{\label{fig:hyb} Scaling of the one-loop fermion self-energy  $\Im \Sigma(\om, k_\perp)$ along the hybridized quantum critical line in (2+1)-d. $\nu\equiv \nu_B$ denotes the bosonic scaling constant.
 }
\end{center}
\end{figure}

\subsubsection{ $\nu_B > 1$ }

The imaginary part is approximated by
\be
   \Im \chi(\om, \vec k) \sim {c'' \om^{2\nu_B} \ov \le(c_1 k^2 + c_2 \om^2\ri)^2  }
\ee
%where we neglected an $\om^{2\nu_B}$ term from the denominator.

We obtain the following limits,
\be
\Im \Sigma \sim  \left\{ \begin{array}{ll}
 {c'' \pi  \ov 2 c_1^2 (1+ 2\nu_B) }  \cdot{\Om^{2\nu_B+1} \ov |k_\perp|^3}
%{c_i  c_2^{3/ 2} \pi v_F^3 \ov 2  (2\nu_B+1) c_1^{7/ 2}}  {\Om^{2\nu_B+1} \ov |k_\perp|^3}
& {\rm for} \quad {0 < \Om \ll  |v_F k_\perp|}
 \\
 & \\
{c'' \pi  v_F^3 \ov  4 c_1^2 \le(1+ { c_2 v_F^2 \ov c_1} \ri)^{3\ov 2}} \cdot{\Om^{2\nu_B-2}  \ov \nu_B-1}  & {\rm for} \quad {\Om =  |v_F k_\perp|}
\\
 & \\
{c'' \pi  v_F^3 f \ov 2 c_1^2 (1+ 2\nu_B) } \cdot{\Om^{2\nu_B-2} }  &  \hskip-0.0cm {\rm for} \ \
%{\Om_{\rm UV} \gg \Om \gg  |v_F k_\perp|}
{|v_F k_\perp| \ll  \Om  %\ll \Om_{\rm UV}
 }
\end{array} \right.
\ee
{\footnotesize
\be
 \nonumber
  f =  F_1\le(1+2\nu_B, {3\ov 2},{3\ov 2},2+2\nu_B; \, 1+i\sqrt{c_2 \ov c_1} v_F,1-i\sqrt{c_2 \ov c_1} v_F \ri)
\ee
}
where $F_1$ is the Appell hypergeometric function of two variables.
There is a crossover at the energy scale set by $v_F k_\perp$. Above this scale, the self-energy becomes independent of the momentum.

\subsubsection{ $ \nu_B = 1$ }

Let us neglect logarithms and approximate the susceptibility by the simple function,
\be
   \chi(\om, \vec k) \sim {1 \ov c_1 k^2 +  c \om^{2} }
\ee
The fermion self-energy is then a scaling function
$
  \Im \Sigma(\Om, k) = f\le({\Om \ov v_F k_\perp}\ri)
$.
In the two limits,
\be
\Im \Sigma \sim  \left\{ \begin{array}{ll}
 {\pi c''  \ov 6 c_1^2} {\Om^{3} \ov |k_\perp|^3} & {\rm for} \quad {|\Om| \ll  |v_F k_\perp|} \\
 & \\
C & {\rm for} \quad {|\Om| \gg  |v_F k_\perp|}
\end{array} \right.
\ee
where the constant is
{\footnotesize
\be
\nonumber
 C =  {{c''} \pi  \alpha^{5\ov 2}  \le( (\sqrt{\alpha} - 1)\sqrt{1+\alpha} + \alpha \log\le[ \le(\sqrt{1+{1\ov \alpha}}-1 \ri)\le(\sqrt{1+\alpha}-1  \ri)  \ri]   \ri) \ov 2 c_2^2 v_F (1+\alpha)^{3\ov 2} }
\ee
}
with $\alpha =    {c_2 v_F^2 \ov c_1 }$.

\subsubsection{  $0< \nu_B < 1$ }

The imaginary part of the susceptibility can be approximated by
\be
   \Im \chi(\om, \vec k) \sim {  c'' \om^{2\nu_B}  \ov \le(c_1 k^2 +   c' \om^{2\nu_B} \ri)^2  + c''^2\om^{4\nu_B}  }
\ee
The transverse momentum integration is straightforward, and the frequency integral can be done in two limits\footnote{ This form of the self-energy has been confirmed by numerical calculations and an analytical calculation at $\nu_B=\half$ where the integrals can be evaluated exactly.}${}^{,}$\footnote{We found the same result in \cite{wolf}; see eqn. 136 on page 1040.}
\be
\Im \Sigma \sim  \left\{ \begin{array}{ll}
{c'' \pi  \ov  2   c_1^2(1+ 2\nu_B)} \cdot {\Om^{2\nu_B+1} \ov |k_\perp|^3} & {\rm for} \quad {|\Om| \ll  \alpha|k_\perp|^{1/\nu_B}} \\
 & \\
 { \pi  \sin {\varphi\ov 2} \ov   \sqrt{c_1 |c|}} \cdot{\Om^{1-\nu_B}\ov  1-\nu_B } & \hskip-0.0cm {\rm for} \quad {  \alpha|k_\perp|^{1/\nu_B} \ll |\Om|  %\ll \tilde\Om_{\rm UV}
 }
\end{array} \right.
\ee
where $\alpha = |{c_1 / c}|^{1 \ov 2\nu_B}$ and $c = |c|e^{i\varphi}$. Near the marginal critical point ($\nu_B \to 0$), the small frequency region rapidly disappears and the self-energy becomes momentum-independent. Note also that the low-frequency regime has the same form as that of the $\nu_B >1 $ case.

\subsection{Hybridized critical point in 3d }

Here we assume that the results of Section III apply to 3+1 boundary spacetime dimensions.
Then, similarly to the (2+1)-dimensional case, we can write
{
\bea
  &&  \Im  \Sigma(k_\perp, \Om) \sim  {\rm sgn \, \Om } \, \Lambda(k_F)^2 \int_{0}^{\Om} {d\om_1 \over 2\pi} \\
  \nonumber
  &&  \hskip -0.3cm  \times  \int_{0}^{\infty} dx \,
    \le. 2\pi x \, \Im \chi(\om, q) \ri.\bigg|_{\om = \Om-\om_1; \ q = \sqrt{\le(k_\perp - {\om_1 \ov v_F} \ri)^2 + x^2}}
\eea
}
Compared to the 2d case, the integrand has a $2\pi x$ factor which comes from the momentum integral in the extra dimension.

\begin{figure}[h] \begin{center}
\includegraphics[scale=0.8]{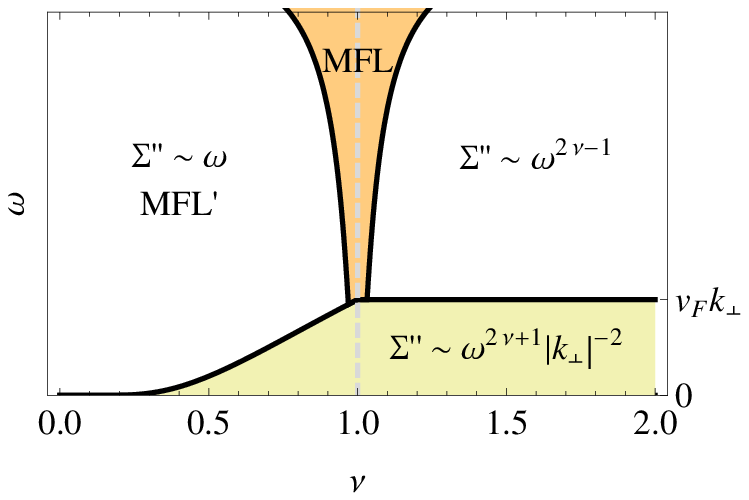}
\caption{\label{fig:hyb3d} Scaling of the one-loop fermion self-energy $\Im \Sigma(\om, k_\perp)$ along the hybridized quantum critical line in (3+1)-d. $\nu\equiv \nu_B$ denotes the bosonic scaling constant.}
\end{center}
\end{figure}

\subsubsection{ $\nu_B > 1$ }

We have,
\be
\Im \Sigma \sim  \left\{ \begin{array}{ll}
 {c'' \pi  \ov c_1^2 (1+ 2\nu_B) }  \cdot{\Om^{2\nu_B+1} \ov |k_\perp|^2}
%{c_i  c_2^{3/ 2} \pi v_F^3 \ov 2  (2\nu_B+1) c_1^{7/ 2}}  {\Om^{2\nu_B+1} \ov |k_\perp|^3}
& {\rm for} \quad {0 < \Om \ll  |v_F k_\perp|}
 \\
 & \\
{c'' \pi  v_F^2 \ov   c_1^2 \le(1+ { c_2 v_F^2 \ov c_1} \ri) } \cdot{\Om^{2\nu_B-1}  \ov 2\nu_B-1}  & {\rm for} \quad {\Om =  |v_F k_\perp|}
\\
 & \\
{c'' \pi  v_F \tilde{f} \ov  c_1^{3/2} c_2^{1/2}   } \cdot{\Om^{2\nu_B-1} }  &  \hskip-0.5cm {\rm for} \ \
{|v_F k_\perp| \ll  \Om \ll \Om_{\rm UV}   }
\end{array} \right.
\ee
where $\tilde{f}$ can be expressed in terms of a hypergeometric function of $\nu_B, c_{1,2}$ and $v_F$.

\subsubsection{ $ \nu_B = 1$ }

The self-energy,
\be
\Im \Sigma \sim  \left\{ \begin{array}{ll}
 { c'' \pi \ov 3 c_1^2} {\Om^{3} \ov |k_\perp|^2} & {\rm for} \quad {|\Om| \ll  |v_F k_\perp|} \\
 & \\
{\tilde{C} \ov c_1}  \Om & {\rm for} \quad {|\Om| \gg  |v_F k_\perp|}
\end{array} \right.
\ee
where the constant is
{%\footnotesize
\be
\nonumber
 \tilde{C} =  {\pi^2 \ov 2 } - {\pi  } \int_0^1 dw \,\arctan { c_2 + c' + {c_1 w^2 \ov v_F^2 (w-1)^2} \ov c'' }
\ee
}

\subsubsection{  $0< \nu_B < 1$ }

Here we have,
\be
\Im \Sigma \sim  \left\{ \begin{array}{ll}
{ c'' \pi  \ov   c_1^2(1+ 2\nu_B)} \cdot {\Om^{2\nu_B+1} \ov |k_\perp|^2} & {\rm for} \quad {|\Om| \ll  \alpha|k_\perp|^{1/\nu_B}} \\
 & \\
 { \pi \le( {\pi \ov 2} -  \arctan {c' \ov c''} \ri)\Om \ov   c_1 }      & \hskip-0.2cm {\rm for} \ \ {  \alpha|k_\perp|^{1/\nu_B} \ll |\Om|  \ll \tilde\Om_{\rm UV}}
\end{array} \right.
\ee

\subsection{Away from the critical point}

In the bifurcating case, the boson spectral function has the frequency dependence,
\be
   \chi(  \om) \sim  c \om^{2\nu_B}
\ee
where $c\in \field{C}$. We neglect the momentum dependence of $\nu_B$. Then,
\be
 \nonumber
  \Im  \Sigma(\Om) \sim \int_{0}^{\Om} {d\om_1 \over 2\pi}
      \Im c \le(\Om - \om_1  \ri)^{2\nu_B} \sim  \Im\le( c \, \Om^{2\nu_B+1} \ri)
\ee
When $\nu_B^2 < 0$, this locks the frequency scaling to be linear. Although the formula suggests a log-oscillating self-energy, it can only be trusted at high enough frequencies where the first oscillation has not happened yet. Below this energy scale, the scalar soliton kills the oscillating behavior.

Near the hybridized critical line,
\be
   \chi(\om, \vec k) \sim {1 \ov \delta + c_1 k^2 + c_2 \om^2 + c(k) \om^{2\nu_B} }
\ee
where $\delta$ parametrizes the distance to the QCP. The result for $\nu_B > 1$ is
\be
 \nonumber
  \Im  \Sigma(\Om, k) \sim  { \Im\le( c \, \Om^{2\nu_B+1} \ri) \ov \le[\delta+ \max({\Om \over v_F}, k_\perp)^2 \ri]^{3\over 2}}
\ee

\section{Discussion}

In this paper, we studied the holographic response function of fermion operators at various quantum phase transitions. The order parameter is a neutral scalar field in the bulk which may represent antiferromagnetic order.
The phase diagram is shown in FIG. \ref{fig:phase}.

When the scalar $m^2$ reaches the $AdS_2$ Breitenlohner-Freedman bound at the `bifurcating' critical line, a classical instability develops and the system enters the ordered phase.  Certain bulk loop corrections to the boundary fermionic two-point functions can be reduced to ordinary field theory calculations where the vertices are replaced by effective interaction vertices.
These effective vertices are determined by radial integrals in the bulk. At the `marginal' critical point, the one-loop bulk contribution from the scalar to the fermionic self-energy is $\Sigma_{(1)} \sim  \om \log \om$. If the absence of other relevant low-energy effects, the Green's function can be written as
 \be
  {G}(\om, k) = {h_1 \over (k-k_F) - {\om \over v_F} + \Sigma_{(0)}  + N^{-2}\Sigma_{(1)}  }
\ee
which would give a marginal Fermi liquid.
Note that bulk loop diagrams are suppressed by powers of $N$ and the one-loop results can only be important at very small energy scales. In the case of a marginal Fermi liquid, this may happen at exponentially low energies where the logarithm cancels the $N^{-2}$ factor. However, at such energy scales it is important to take into account the backreaction of the bulk charge density on the background geometry. This deforms the near-horizon $AdS_2$ into a Lifshitz space \cite{Hartnoll:2009ns, Hartnoll:2010gu, Hartnoll:2011dm, Iqbal:2011in, Cubrovic:2011xm} and may invalidate the one-loop result.

Double trace operators can be used to drive the system into the symmetry breaking phase through the `hybridized' quantum critical point. The one-loop results in two spatial boundary dimensions give $\Sigma_{(1)} \sim  \om^{1-\nu_B}$. Here $\nu_B(m^2)$ is a bosonic scaling exponent which vanishes at the marginal critical point.
In this case, it is important to take into account the Landau damping of the bosons. This presumably changes the low-energy physics in important ways.
% (see \cite{PhysRevB.80.165102} for a

The effective Yukawa vertex in this paper was finite. The finiteness crucially depends on the spinor index structure. It would be interesting to study other bulk interactions where the effective vertex has an infrared divergence (see also \cite{Hung:2010pe}).

Finally, it would be interesting to study a similar system using a consistent `top-down' approach where the boundary field theory and the operators corresponding to the bulk fields are explicitly known.

\vspace{0.2in}   \centerline{\bf{Acknowledgments}} \vspace{0.2in} I thank Thomas Faulkner, Sean Hartnoll, Diego Hofman, Nabil Iqbal, Hong Liu, John McGreevy, Max Metlitski, M\'ark Mezei, and Michael Mulligan for interesting discussions and collaboration on related projects.
I thank the University of Michigan, Stanford University, the Aspen Center for Physics, the Lorentz Center, and the Kavli Institute for Theoretical Physics for hospitality. The author is supported by the DOE grant DE-FG02-92ER40697.

\bibliography{marginal}

\begin{thebibliography}{35}
\expandafter\ifx\csname natexlab\endcsname\relax\def\natexlab#1{#1}\fi
\expandafter\ifx\csname bibnamefont\endcsname\relax
  \def\bibnamefont#1{#1}\fi
\expandafter\ifx\csname bibfnamefont\endcsname\relax
  \def\bibfnamefont#1{#1}\fi
\expandafter\ifx\csname citenamefont\endcsname\relax
  \def\citenamefont#1{#1}\fi
\expandafter\ifx\csname url\endcsname\relax
  \def\url#1{\texttt{#1}}\fi
\expandafter\ifx\csname urlprefix\endcsname\relax\def\urlprefix{URL }\fi
\providecommand{\bibinfo}[2]{#2}
\providecommand{\eprint}[2][]{\url{#2}}

\bibitem[{\citenamefont{{J. G. Bednorz and K. A. M\"uller}}(1986)}]{Bednorz}
\bibinfo{author}{\bibnamefont{{J. G. Bednorz and K. A. M\"uller}}},
  \bibinfo{journal}{Z. Phys.} \textbf{\bibinfo{volume}{B64}},
  \bibinfo{pages}{189} (\bibinfo{year}{1986}).

\bibitem[{\citenamefont{Maldacena}(1998)}]{Maldacena:1997re}
\bibinfo{author}{\bibfnamefont{J.~M.} \bibnamefont{Maldacena}},
  \bibinfo{journal}{Adv. Theor. Math. Phys.} \textbf{\bibinfo{volume}{2}},
  \bibinfo{pages}{231} (\bibinfo{year}{1998}).

\bibitem[{\citenamefont{Gubser et~al.}(1998)\citenamefont{Gubser, Klebanov, and
  Polyakov}}]{Gubser:1998bc}
\bibinfo{author}{\bibfnamefont{S.~S.} \bibnamefont{Gubser}},
  \bibinfo{author}{\bibfnamefont{I.~R.} \bibnamefont{Klebanov}},
  \bibnamefont{and} \bibinfo{author}{\bibfnamefont{A.~M.}
  \bibnamefont{Polyakov}}, \bibinfo{journal}{Phys. Lett.}
  \textbf{\bibinfo{volume}{B428}}, \bibinfo{pages}{105} (\bibinfo{year}{1998}),
  \eprint{hep-th/9802109}.

\bibitem[{\citenamefont{Witten}(1998)}]{Witten:1998qj}
\bibinfo{author}{\bibfnamefont{E.}~\bibnamefont{Witten}},
  \bibinfo{journal}{Adv. Theor. Math. Phys.} \textbf{\bibinfo{volume}{2}},
  \bibinfo{pages}{253} (\bibinfo{year}{1998}).

\bibitem[{\citenamefont{Lee}(2008)}]{Lee:2008xf}
\bibinfo{author}{\bibfnamefont{S.-S.} \bibnamefont{Lee}}
  (\bibinfo{year}{2008}), \eprint{0809.3402}.

\bibitem[{\citenamefont{Liu et~al.}(2009)\citenamefont{Liu, McGreevy, and
  Vegh}}]{Liu:2009dm}
\bibinfo{author}{\bibfnamefont{H.}~\bibnamefont{Liu}},
  \bibinfo{author}{\bibfnamefont{J.}~\bibnamefont{McGreevy}}, \bibnamefont{and}
  \bibinfo{author}{\bibfnamefont{D.}~\bibnamefont{Vegh}}
  (\bibinfo{year}{2009}), \eprint{0903.2477}.

\bibitem[{\citenamefont{Cubrovic et~al.}(2009)\citenamefont{Cubrovic, Zaanen,
  and Schalm}}]{Cubrovic:2009ye}
\bibinfo{author}{\bibfnamefont{M.}~\bibnamefont{Cubrovic}},
  \bibinfo{author}{\bibfnamefont{J.}~\bibnamefont{Zaanen}}, \bibnamefont{and}
  \bibinfo{author}{\bibfnamefont{K.}~\bibnamefont{Schalm}},
  \bibinfo{journal}{Science} \textbf{\bibinfo{volume}{325}},
  \bibinfo{pages}{439} (\bibinfo{year}{2009}), \eprint{0904.1993}.

\bibitem[{\citenamefont{Faulkner et~al.}(2009)\citenamefont{Faulkner, Liu,
  McGreevy, and Vegh}}]{Faulkner:2009wj}
\bibinfo{author}{\bibfnamefont{T.}~\bibnamefont{Faulkner}},
  \bibinfo{author}{\bibfnamefont{H.}~\bibnamefont{Liu}},
  \bibinfo{author}{\bibfnamefont{J.}~\bibnamefont{McGreevy}}, \bibnamefont{and}
  \bibinfo{author}{\bibfnamefont{D.}~\bibnamefont{Vegh}}
  (\bibinfo{year}{2009}), \eprint{0907.2694}.

\bibitem[{\citenamefont{Denef et~al.}(2010)\citenamefont{Denef, Hartnoll, and
  Sachdev}}]{Denef:2009kn}
\bibinfo{author}{\bibfnamefont{F.}~\bibnamefont{Denef}},
  \bibinfo{author}{\bibfnamefont{S.~A.} \bibnamefont{Hartnoll}},
  \bibnamefont{and} \bibinfo{author}{\bibfnamefont{S.}~\bibnamefont{Sachdev}},
  \bibinfo{journal}{Class. Quant. Grav.} \textbf{\bibinfo{volume}{27}},
  \bibinfo{pages}{125001} (\bibinfo{year}{2010}), \eprint{0908.2657}.

\bibitem[{\citenamefont{Faulkner and Polchinski}(2010)}]{Faulkner:2010tq}
\bibinfo{author}{\bibfnamefont{T.}~\bibnamefont{Faulkner}} \bibnamefont{and}
  \bibinfo{author}{\bibfnamefont{J.}~\bibnamefont{Polchinski}}
  (\bibinfo{year}{2010}), \eprint{1001.5049}.

\bibitem[{\citenamefont{Gauntlett et~al.}(2011)\citenamefont{Gauntlett, Sonner,
  and Waldram}}]{Gauntlett:2011mf}
\bibinfo{author}{\bibfnamefont{J.~P.} \bibnamefont{Gauntlett}},
  \bibinfo{author}{\bibfnamefont{J.}~\bibnamefont{Sonner}}, \bibnamefont{and}
  \bibinfo{author}{\bibfnamefont{D.}~\bibnamefont{Waldram}},
  \bibinfo{journal}{Phys.Rev.Lett.} \textbf{\bibinfo{volume}{107}},
  \bibinfo{pages}{241601} (\bibinfo{year}{2011}), \eprint{1106.4694}.

\bibitem[{\citenamefont{Belliard et~al.}(2011)\citenamefont{Belliard, Gubser,
  and Yarom}}]{Belliard:2011qq}
\bibinfo{author}{\bibfnamefont{R.}~\bibnamefont{Belliard}},
  \bibinfo{author}{\bibfnamefont{S.~S.} \bibnamefont{Gubser}},
  \bibnamefont{and} \bibinfo{author}{\bibfnamefont{A.}~\bibnamefont{Yarom}},
  \bibinfo{journal}{JHEP} \textbf{\bibinfo{volume}{1110}}, \bibinfo{pages}{055}
  (\bibinfo{year}{2011}), \eprint{1106.6030}.

\bibitem[{\citenamefont{Jensen et~al.}(2011)\citenamefont{Jensen, Kachru,
  Karch, Polchinski, and Silverstein}}]{Jensen:2011su}
\bibinfo{author}{\bibfnamefont{K.}~\bibnamefont{Jensen}},
  \bibinfo{author}{\bibfnamefont{S.}~\bibnamefont{Kachru}},
  \bibinfo{author}{\bibfnamefont{A.}~\bibnamefont{Karch}},
  \bibinfo{author}{\bibfnamefont{J.}~\bibnamefont{Polchinski}},
  \bibnamefont{and}
  \bibinfo{author}{\bibfnamefont{E.}~\bibnamefont{Silverstein}},
  \bibinfo{journal}{Phys.Rev.} \textbf{\bibinfo{volume}{D84}},
  \bibinfo{pages}{126002} (\bibinfo{year}{2011}), \eprint{1105.1772}.

\bibitem[{\citenamefont{Iqbal et~al.}(2010)\citenamefont{Iqbal, Liu, Mezei, and
  Si}}]{Iqbal:2010eh}
\bibinfo{author}{\bibfnamefont{N.}~\bibnamefont{Iqbal}},
  \bibinfo{author}{\bibfnamefont{H.}~\bibnamefont{Liu}},
  \bibinfo{author}{\bibfnamefont{M.}~\bibnamefont{Mezei}}, \bibnamefont{and}
  \bibinfo{author}{\bibfnamefont{Q.}~\bibnamefont{Si}},
  \bibinfo{journal}{Phys.Rev.} \textbf{\bibinfo{volume}{D82}},
  \bibinfo{pages}{045002} (\bibinfo{year}{2010}), \eprint{1003.0010}.

\bibitem[{\citenamefont{Romans}(1992)}]{Romans:1991nq}
\bibinfo{author}{\bibfnamefont{L.~J.} \bibnamefont{Romans}},
  \bibinfo{journal}{Nucl. Phys.} \textbf{\bibinfo{volume}{B383}},
  \bibinfo{pages}{395} (\bibinfo{year}{1992}), \eprint{hep-th/9203018}.

\bibitem[{\citenamefont{Chamblin et~al.}(1999)\citenamefont{Chamblin, Emparan,
  Johnson, and Myers}}]{Chamblin:1999tk}
\bibinfo{author}{\bibfnamefont{A.}~\bibnamefont{Chamblin}},
  \bibinfo{author}{\bibfnamefont{R.}~\bibnamefont{Emparan}},
  \bibinfo{author}{\bibfnamefont{C.~V.} \bibnamefont{Johnson}},
  \bibnamefont{and} \bibinfo{author}{\bibfnamefont{R.~C.} \bibnamefont{Myers}},
  \bibinfo{journal}{Phys. Rev.} \textbf{\bibinfo{volume}{D60}},
  \bibinfo{pages}{064018} (\bibinfo{year}{1999}), \eprint{hep-th/9902170}.

\bibitem[{\citenamefont{Henningson and Sfetsos}(1998)}]{Henningson:1998cd}
\bibinfo{author}{\bibfnamefont{M.}~\bibnamefont{Henningson}} \bibnamefont{and}
  \bibinfo{author}{\bibfnamefont{K.}~\bibnamefont{Sfetsos}},
  \bibinfo{journal}{Phys. Lett.} \textbf{\bibinfo{volume}{B431}},
  \bibinfo{pages}{63} (\bibinfo{year}{1998}), \eprint{hep-th/9803251}.

\bibitem[{\citenamefont{Mueck and Viswanathan}(1998)}]{Mueck:1998iz}
\bibinfo{author}{\bibfnamefont{W.}~\bibnamefont{Mueck}} \bibnamefont{and}
  \bibinfo{author}{\bibfnamefont{K.~S.} \bibnamefont{Viswanathan}},
  \bibinfo{journal}{Phys. Rev.} \textbf{\bibinfo{volume}{D58}},
  \bibinfo{pages}{106006} (\bibinfo{year}{1998}), \eprint{hep-th/9805145}.

\bibitem[{\citenamefont{Iqbal and Liu}(2009)}]{Iqbal:2009fd}
\bibinfo{author}{\bibfnamefont{N.}~\bibnamefont{Iqbal}} \bibnamefont{and}
  \bibinfo{author}{\bibfnamefont{H.}~\bibnamefont{Liu}},
  \bibinfo{journal}{Fortsch. Phys.} \textbf{\bibinfo{volume}{57}},
  \bibinfo{pages}{367} (\bibinfo{year}{2009}).

\bibitem[{\citenamefont{Faulkner
  et~al.}(2010{\natexlab{a}})\citenamefont{Faulkner, Horowitz, McGreevy,
  Roberts, and Vegh}}]{Faulkner:2009am}
\bibinfo{author}{\bibfnamefont{T.}~\bibnamefont{Faulkner}},
  \bibinfo{author}{\bibfnamefont{G.~T.} \bibnamefont{Horowitz}},
  \bibinfo{author}{\bibfnamefont{J.}~\bibnamefont{McGreevy}},
  \bibinfo{author}{\bibfnamefont{M.~M.} \bibnamefont{Roberts}},
  \bibnamefont{and} \bibinfo{author}{\bibfnamefont{D.}~\bibnamefont{Vegh}},
  \bibinfo{journal}{JHEP} \textbf{\bibinfo{volume}{03}}, \bibinfo{pages}{121}
  (\bibinfo{year}{2010}{\natexlab{a}}), \eprint{0911.3402}.

\bibitem[{\citenamefont{Faulkner
  et~al.}(2011{\natexlab{a}})\citenamefont{Faulkner, Horowitz, and
  Roberts}}]{Faulkner:2010gj}
\bibinfo{author}{\bibfnamefont{T.}~\bibnamefont{Faulkner}},
  \bibinfo{author}{\bibfnamefont{G.~T.} \bibnamefont{Horowitz}},
  \bibnamefont{and} \bibinfo{author}{\bibfnamefont{M.~M.}
  \bibnamefont{Roberts}}, \bibinfo{journal}{JHEP}
  \textbf{\bibinfo{volume}{1104}}, \bibinfo{pages}{051}
  (\bibinfo{year}{2011}{\natexlab{a}}), \eprint{1008.1581}.

\bibitem[{\citenamefont{Iqbal et~al.}(2011{\natexlab{a}})\citenamefont{Iqbal,
  Liu, and Mezei}}]{Iqbal:2011aj}
\bibinfo{author}{\bibfnamefont{N.}~\bibnamefont{Iqbal}},
  \bibinfo{author}{\bibfnamefont{H.}~\bibnamefont{Liu}}, \bibnamefont{and}
  \bibinfo{author}{\bibfnamefont{M.}~\bibnamefont{Mezei}}
  (\bibinfo{year}{2011}{\natexlab{a}}), \eprint{1108.0425}.

\bibitem[{\citenamefont{Varma et~al.}(1989)\citenamefont{Varma, Littlewood,
  Schmitt-Rink, Abrahams, and Ruckenstein}}]{varma}
\bibinfo{author}{\bibfnamefont{C.~M.} \bibnamefont{Varma}},
  \bibinfo{author}{\bibfnamefont{P.~B.} \bibnamefont{Littlewood}},
  \bibinfo{author}{\bibfnamefont{S.}~\bibnamefont{Schmitt-Rink}},
  \bibinfo{author}{\bibfnamefont{E.}~\bibnamefont{Abrahams}}, \bibnamefont{and}
  \bibinfo{author}{\bibfnamefont{A.~E.} \bibnamefont{Ruckenstein}},
  \bibinfo{journal}{Phys. Rev. Lett.} \textbf{\bibinfo{volume}{63}},
  \bibinfo{pages}{1996} (\bibinfo{year}{1989}).

\bibitem[{\citenamefont{Jensen}(2011)}]{Jensen:2011af}
\bibinfo{author}{\bibfnamefont{K.}~\bibnamefont{Jensen}},
  \bibinfo{journal}{Phys.Rev.Lett.} \textbf{\bibinfo{volume}{107}},
  \bibinfo{pages}{231601} (\bibinfo{year}{2011}), \eprint{1108.0421}.

\bibitem[{\citenamefont{Georges et~al.}(2001)\citenamefont{Georges, Parcollet,
  and Sachdev}}]{sachdevt}
\bibinfo{author}{\bibfnamefont{A.}~\bibnamefont{Georges}},
  \bibinfo{author}{\bibfnamefont{O.}~\bibnamefont{Parcollet}},
  \bibnamefont{and} \bibinfo{author}{\bibfnamefont{S.}~\bibnamefont{Sachdev}},
  \bibinfo{journal}{Phys. Rev. B} \textbf{\bibinfo{volume}{63}},
  \bibinfo{pages}{134406} (\bibinfo{year}{2001}),
  \urlprefix\url{http://link.aps.org/doi/10.1103/PhysRevB.63.134406}.

\bibitem[{\citenamefont{Faulkner et~al.}()\citenamefont{Faulkner, Iqbal, Liu,
  McGreevy, and Vegh}}]{conduc}
\bibinfo{author}{\bibfnamefont{T.}~\bibnamefont{Faulkner}},
  \bibinfo{author}{\bibfnamefont{N.}~\bibnamefont{Iqbal}},
  \bibinfo{author}{\bibfnamefont{H.}~\bibnamefont{Liu}},
  \bibinfo{author}{\bibfnamefont{J.}~\bibnamefont{McGreevy}}, \bibnamefont{and}
  \bibinfo{author}{\bibfnamefont{D.}~\bibnamefont{Vegh}},
  \emph{\bibinfo{title}{{Charge transport by holographic Fermi surfaces. Work
  in progress}}}.

\bibitem[{\citenamefont{Faulkner
  et~al.}(2010{\natexlab{b}})\citenamefont{Faulkner, Iqbal, Liu, McGreevy, and
  Vegh}}]{Faulkner:2010zz}
\bibinfo{author}{\bibfnamefont{T.}~\bibnamefont{Faulkner}},
  \bibinfo{author}{\bibfnamefont{N.}~\bibnamefont{Iqbal}},
  \bibinfo{author}{\bibfnamefont{H.}~\bibnamefont{Liu}},
  \bibinfo{author}{\bibfnamefont{J.}~\bibnamefont{McGreevy}}, \bibnamefont{and}
  \bibinfo{author}{\bibfnamefont{D.}~\bibnamefont{Vegh}},
  \bibinfo{journal}{Science} \textbf{\bibinfo{volume}{329}},
  \bibinfo{pages}{1043} (\bibinfo{year}{2010}{\natexlab{b}}).

\bibitem[{\citenamefont{Hung and Shang}(2011)}]{Hung:2010pe}
\bibinfo{author}{\bibfnamefont{L.-Y.} \bibnamefont{Hung}} \bibnamefont{and}
  \bibinfo{author}{\bibfnamefont{Y.}~\bibnamefont{Shang}},
  \bibinfo{journal}{Phys.Rev.} \textbf{\bibinfo{volume}{D83}},
  \bibinfo{pages}{024029} (\bibinfo{year}{2011}), \eprint{1007.2653}.

\bibitem[{\citenamefont{Faulkner
  et~al.}(2011{\natexlab{b}})\citenamefont{Faulkner, Iqbal, Liu, McGreevy, and
  Vegh}}]{Faulkner:2011tm}
\bibinfo{author}{\bibfnamefont{T.}~\bibnamefont{Faulkner}},
  \bibinfo{author}{\bibfnamefont{N.}~\bibnamefont{Iqbal}},
  \bibinfo{author}{\bibfnamefont{H.}~\bibnamefont{Liu}},
  \bibinfo{author}{\bibfnamefont{J.}~\bibnamefont{McGreevy}}, \bibnamefont{and}
  \bibinfo{author}{\bibfnamefont{D.}~\bibnamefont{Vegh}}
  (\bibinfo{year}{2011}{\natexlab{b}}), \eprint{1101.0597}.

\bibitem[{\citenamefont{L\"ohneysen et~al.}(2007)\citenamefont{L\"ohneysen,
  Rosch, Vojta, and W\"olfle}}]{wolf}
\bibinfo{author}{\bibfnamefont{H.~v.} \bibnamefont{L\"ohneysen}},
  \bibinfo{author}{\bibfnamefont{A.}~\bibnamefont{Rosch}},
  \bibinfo{author}{\bibfnamefont{M.}~\bibnamefont{Vojta}}, \bibnamefont{and}
  \bibinfo{author}{\bibfnamefont{P.}~\bibnamefont{W\"olfle}},
  \bibinfo{journal}{Rev. Mod. Phys.}  (\bibinfo{year}{2007}).

\bibitem[{\citenamefont{Hartnoll et~al.}(2010)\citenamefont{Hartnoll,
  Polchinski, Silverstein, and Tong}}]{Hartnoll:2009ns}
\bibinfo{author}{\bibfnamefont{S.~A.} \bibnamefont{Hartnoll}},
  \bibinfo{author}{\bibfnamefont{J.}~\bibnamefont{Polchinski}},
  \bibinfo{author}{\bibfnamefont{E.}~\bibnamefont{Silverstein}},
  \bibnamefont{and} \bibinfo{author}{\bibfnamefont{D.}~\bibnamefont{Tong}},
  \bibinfo{journal}{JHEP} \textbf{\bibinfo{volume}{1004}}, \bibinfo{pages}{120}
  (\bibinfo{year}{2010}), \eprint{0912.1061}.

\bibitem[{\citenamefont{Hartnoll and Tavanfar}(2011)}]{Hartnoll:2010gu}
\bibinfo{author}{\bibfnamefont{S.~A.} \bibnamefont{Hartnoll}} \bibnamefont{and}
  \bibinfo{author}{\bibfnamefont{A.}~\bibnamefont{Tavanfar}},
  \bibinfo{journal}{Phys.Rev.} \textbf{\bibinfo{volume}{D83}},
  \bibinfo{pages}{046003} (\bibinfo{year}{2011}), \eprint{1008.2828}.

\bibitem[{\citenamefont{Hartnoll et~al.}(2011)\citenamefont{Hartnoll, Hofman,
  and Vegh}}]{Hartnoll:2011dm}
\bibinfo{author}{\bibfnamefont{S.~A.} \bibnamefont{Hartnoll}},
  \bibinfo{author}{\bibfnamefont{D.~M.} \bibnamefont{Hofman}},
  \bibnamefont{and} \bibinfo{author}{\bibfnamefont{D.}~\bibnamefont{Vegh}},
  \bibinfo{journal}{JHEP} \textbf{\bibinfo{volume}{1108}}, \bibinfo{pages}{096}
  (\bibinfo{year}{2011}), \eprint{1105.3197}.

\bibitem[{\citenamefont{Iqbal et~al.}(2011{\natexlab{b}})\citenamefont{Iqbal,
  Liu, and Mezei}}]{Iqbal:2011in}
\bibinfo{author}{\bibfnamefont{N.}~\bibnamefont{Iqbal}},
  \bibinfo{author}{\bibfnamefont{H.}~\bibnamefont{Liu}}, \bibnamefont{and}
  \bibinfo{author}{\bibfnamefont{M.}~\bibnamefont{Mezei}}
  (\bibinfo{year}{2011}{\natexlab{b}}), \eprint{1105.4621}.

\bibitem[{\citenamefont{Cubrovic et~al.}(2011)\citenamefont{Cubrovic, Liu,
  Schalm, Sun, and Zaanen}}]{Cubrovic:2011xm}
\bibinfo{author}{\bibfnamefont{M.}~\bibnamefont{Cubrovic}},
  \bibinfo{author}{\bibfnamefont{Y.}~\bibnamefont{Liu}},
  \bibinfo{author}{\bibfnamefont{K.}~\bibnamefont{Schalm}},
  \bibinfo{author}{\bibfnamefont{Y.-W.} \bibnamefont{Sun}}, \bibnamefont{and}
  \bibinfo{author}{\bibfnamefont{J.}~\bibnamefont{Zaanen}},
  \bibinfo{journal}{Phys.Rev.} \textbf{\bibinfo{volume}{D84}},
  \bibinfo{pages}{086002} (\bibinfo{year}{2011}), \eprint{1106.1798}.

\end{thebibliography}

\end{document}